\documentclass[
 aip,
 jcp,
 amsmath,amssymb,
 reprint,
]{revtex4-1}

\usepackage{graphicx}
\usepackage{dcolumn}
\usepackage{bm}
\usepackage[utf8]{inputenc}
\usepackage[T1]{fontenc}
\usepackage{mathptmx}
\usepackage{etoolbox}
\usepackage{mathtools}
\usepackage{mathrsfs}
\usepackage{amsthm}
\usepackage{physics}
\usepackage{xcolor}
\usepackage{enumitem}
\usepackage{microtype}
\usepackage{hyperref}

\allowdisplaybreaks

\makeatletter
\def\@email#1#2{%
 \endgroup
 \patchcmd{\titleblock@produce}
  {\frontmatter@RRAPformat}
  {\frontmatter@RRAPformat{\produce@RRAP{*#1\href{mailto:#2}{#2}}}\frontmatter@RRAPformat}
  {}{}
}%
\makeatother

\newcommand{\R}{\mathbb{R}}

\newcommand{\pair}[2]{\langle #1, #2\rangle}
\newcommand{\Trf}{\operatorname{Tr}}

\newcommand{\TrF}{\operatorname{Tr}_{\mathcal F}}

\newcommand{\Pet}{\mathcal P_{\mathrm{et}}}
\newcommand{\Pdiag}{\mathcal P_{\mathrm d}}
\newcommand{\Iet}{\mathcal I_{\mathrm{et}}}
\newcommand{\Idiag}{\mathcal I_{\mathrm d}}
\newcommand{\Xrho}{X_\rho}
\newcommand{\Xgamma}{X_\gamma}
\newcommand{\XG}{X_G}
\newcommand{\XT}{X_T}
\newcommand{\Xrhobeta}{X_{\rho,\beta}}
\newcommand{\Xrhobetastar}{X_{\rho,\beta}^{\ast}}

\begin{document}

\preprint{AIP/123-QED}

\title[]
{RPA as a Hessian Closure: Effective Functionals and Source--Variable Duality Across DFT, LR-TDDFT, 1RDMFT, and MBPT}

\author{Nan Sheng}
\affiliation{
Institute for Computational and Mathematical Engineering (ICME),
Stanford University, Stanford, CA 94305, USA.
}
\email{nansheng@stanford.edu}

\date{\today}

\begin{abstract}
We present a variational formulation of the random phase approximation (RPA) that places density functional theory (DFT), linear-response time-dependent density functional theory (LR-TDDFT), one-body reduced density matrix functional theory (1RDMFT), and Green's function many-body perturbation theory (MBPT) into a common source--variable hierarchy. The central claim is that a broad class of RPA constructions can be organized, independently of any one problem-specific formula, diagrammatic resummation, or small-amplitude equation of motion, as closure approximations to the exact Hessian of an effective functional. In this language, exact linear response is governed by the Hessian of the corresponding effective functional, while RPA is obtained by retaining a reference contribution together with an explicit interaction kernel and discarding the irreducible remainder.
The hierarchy has two independent enrichments of the density-level description. One may enlarge the static local density to a time-dependent density, giving the dynamical density channel of LR-TDDFT, or enlarge it to an equal-time bilocal one-body reduced density matrix, giving the static bilocal channel of 1RDMFT. The Green's function level combines both enrichments, since the one-particle Green's function is bilocal in both space and time. This picture clarifies the relation between DFT, LR-TDDFT, 1RDMFT, and MBPT through exact forward reductions and source restrictions, while emphasizing that the corresponding RPA closures need not commute under projection. The hierarchy also distinguishes the local branch-wise Legendre geometry common to all four levels from the stronger global convex duality that may emerge only in source sectors possessing additional positivity and global regularity.
\end{abstract}

\newtheorem{definition}{Definition}
\newtheorem{proposition}{Proposition}

\maketitle

\section{Introduction}

The random phase approximation (RPA) appears across many-body physics, electronic structure theory, nuclear theory, and soft matter field theory, but it is not usually introduced through a single formal language.\cite{Ren2012,Co2023,Hedin1965,Baym1962,BaymKadanoff1961,Peotta2022} Depending on context, it is presented as a ring-diagram resummation, a density-response approximation, a small-amplitude mean-field theory, or a screening construction. These formulations are successful within their native settings, but they often leave implicit a basic structural distinction between three ingredients: the choice of basic variable, the exact response theory associated with that variable, and the approximation that closes that response theory. As a result, the term ``RPA'' frequently refers to a family of closely related constructions rather than to one transparently stated variational object.

The central claim of the present paper is that a broad class of RPA constructions is naturally understood as closure approximations applied to the exact Hessian of an effective functional associated with a chosen source--variable duality. In this formulation, exact linear response is governed by the Hessian of the corresponding effective functional, while RPA is obtained by retaining a reference contribution together with an explicit bilinear interaction kernel and discarding the irreducible remainder in the exact Hessian. This point of view is not tied to any one problem-specific formula. Its purpose is to isolate the common variational structure that underlies several standard realizations of RPA.

We develop this perspective at four variational levels organized by two independent enrichments of the density-level description. At the static density level, the source is a local scalar potential and the basic variable is the density.\cite{HohenbergKohn1964,KohnSham1965,Levy1979,Lieb1983,Teale2022} At the dynamical density level, the source is a time-dependent local scalar potential and the response object is the time-dependent density response, giving the standard LR-TDDFT setting.\cite{RungeGross1984,Casida1995,Petersilka1996} At the equal-time bilocal level, the source is a nonlocal one-body potential and the basic variable is the one-body reduced density matrix.\cite{Coleman1963,Gilbert1975,Valone1980,Blochl2013} At the full spacetime-bilocal level, the source is bilocal in space and time and the basic variable is the one-particle Green's function.\cite{Baym1962,BaymKadanoff1961,LuttingerWard1960,Klein1961,CJT1974,Hedin1965,Almbladh1999}

These four levels form a two-directional hierarchy rather than a single linear chain. From static DFT, one may move in a dynamical direction to LR-TDDFT, or in a spatially nonlocal direction to 1RDMFT. MBPT combines both enrichments. Thus the hierarchy contains two natural paths,
\[
\mathrm{DFT}\to \mathrm{1RDMFT}\to \mathrm{MBPT},
\qquad
\mathrm{DFT}\to \mathrm{LR\mbox{-}TDDFT}\to \mathrm{MBPT},
\]
understood as paths of retained information, not as a claim that the associated Hessian closures commute under projection.

The density level provides the sharpest canonical example. There, the exact theory already organizes the inverse density response into a reference contribution, the Coulomb kernel, and an exchange-correlation remainder. Direct density-response RPA is obtained by discarding that remainder.\cite{RungeGross1984,Casida1995,Petersilka1996,FuchsGonze2002,Ren2012,Woods2021} At the static density level this gives the zero-frequency density response and static screening. At the dynamical density level, the same density-channel closure gives the frequency-dependent response used in LR-TDDFT and in the fluctuation-dissipation representation of the adiabatic connection. The equal-time bilocal and full Green's function levels show that this is not a special accident of density functional theory, but the most compressed representative of a broader hierarchy of Hessian closures.

The individual inverse-response identities used below are standard within their respective theories. The contribution of the present work is their organization into a common Hessian-closure framework, the identification of two independent enrichments of the density-level description, and the resulting distinction between exact reduction of response and approximation by closure. In particular, closure and reduction generally do not commute.

The purpose of the present article is therefore not to review applications of RPA, nor to replace the standard formulations used in particular subfields. It is to state clearly the common variational content behind them. Section~\ref{sec:framework} formulates the abstract source--variable framework and the associated Hessian definition of RPA. Section~\ref{sec:levels} develops its realizations in DFT, LR-TDDFT, 1RDMFT, and MBPT, and clarifies the projection and reduction relations between these levels. Section~\ref{sec:discussion} discusses the broader implications of this perspective and the formal place of related RPA-type constructions.

\section{RPA as a Hessian closure in source--variable variational theories}
\label{sec:framework}

We now formulate the abstract variational structure used throughout the paper. The logic of the section is simple. First, one fixes a basic variable and its conjugate source, and defines the corresponding effective functional. Second, one identifies exact linear response with the inverse Hessian of that functional. Third, one defines RPA by specifying a decomposition of the exact Hessian and discarding its irreducible remainder. The purpose of the section is therefore not to introduce a new formalism, but to separate the exact response theory from the closure that turns it into RPA.

\subsection{Source--variable duality and effective functionals}

Fix a real dual pair
\begin{equation}
(X,X^\ast),
\label{eq:dual-pair}
\end{equation}
with canonical pairing \(\pair{J}{q}\). The physical variables need not fill the whole linear space: more generally they lie in a constrained or affine domain
\begin{equation}
\mathcal C\subset q_0+X,
\label{eq:admissible_affine_domain}
\end{equation}
and local variations belong to the appropriate tangent space \(T_q\mathcal C\). Sources belong to the corresponding dual space, modulo any gauge or conserved null directions. The precise choices depend on the theory and will be specified in the concrete realizations below.

We begin from a source-side energy-like functional
\begin{equation}
E[J].
\label{eq:EJ}
\end{equation}
The weakest structure required in this paper is local. On any differentiable source branch \(\mathcal B\) for which
\begin{equation}
q[J]=\frac{\delta E[J]}{\delta J}
\label{eq:q_from_E_general}
\end{equation}
is locally invertible on the responsive tangent space, one may write \(J=J[q]\) and define the local Legendre transform
\begin{equation}
\Gamma_{\mathcal B}[q]
=
E[J[q]]-\pair{J[q]}{q}.
\label{eq:Gamma-local}
\end{equation}
This construction requires neither convexity nor a global extremal principle; the resulting effective functional may be of saddle type. In the differentiable setting, the dual relations are
\begin{equation}
q = \frac{\delta E[J]}{\delta J},
\qquad
-\,J = \frac{\delta \Gamma_{\mathcal B}[q]}{\delta q}.
\label{eq:dual-relations}
\end{equation}

Additional positivity and global regularity strengthen this local geometry. If \(E\) is proper and concave on a convex source domain, with the relevant supremum attained, the same effective functional admits the Legendre--Fenchel representation
\begin{equation}
\Gamma[q]
=
\sup_J
\left\{
E[J]-\pair{J}{q}
\right\}.
\label{eq:Gamma-LF}
\end{equation}
More generally, in this convex setting one may write
\begin{equation}
q \in \partial^+ E[J]
\quad\Longleftrightarrow\quad
-\,J \in \partial \Gamma[q],
\label{eq:subgradient-duality}
\end{equation}
where \(\partial^+\) denotes the concave superdifferential of \(E\), and \(\partial\) the convex subdifferential of \(\Gamma\). Thus global convex duality is an enhanced structure available in suitable theories or source sectors, rather than a prerequisite for the Hessian framework itself.

Equations~\eqref{eq:Gamma-local} and \eqref{eq:Gamma-LF} describe two strengths of the same source--variable geometry. The effective functional is not an additional approximation: it is the exact variable-side object associated locally, and when available globally, with the chosen source-side functional.\cite{Lieb1983,PolonyiSailer2002}

At this stage one must also distinguish the construction of \(\Gamma[q]\) from the physical variational problem. In either the local construction, Eq.~\eqref{eq:Gamma-local}, or the stronger global construction, Eq.~\eqref{eq:Gamma-LF}, the source is varied in order to define the functional itself. The physical state is obtained only after the actual source carried by the problem has been specified. If that source is \(J_{\mathrm{phys}}\), then the physical configuration \(\bar q\) satisfies
\begin{equation}
\frac{\delta \Gamma[\bar q]}{\delta q} = -\,J_{\mathrm{phys}},
\label{eq:EL1}
\end{equation}
or equivalently
\begin{equation}
0=
\frac{\delta}{\delta q}
\Big(
\Gamma[q]+\pair{J_{\mathrm{phys}}}{q}
\Big)\Big|_{q=\bar q}.
\label{eq:EL2}
\end{equation}
If the source used in \(E[J]\) is merely an auxiliary probe and the physical problem corresponds to setting it to zero, then \(J_{\mathrm{phys}}=0\) and one recovers the simpler condition
\[
\frac{\delta \Gamma[\bar q]}{\delta q}=0.
\]

\subsection{Exact response as inverse Hessian}

Once the effective functional has been defined, exact linear response follows directly from the stationarity condition. Linearizing Eq.~\eqref{eq:EL1} around the physical configuration \(\bar q\) gives
\begin{equation}
-\,\delta J
=
\frac{\delta^2 \Gamma[\bar q]}{\delta q\,\delta q}\,
\delta q.
\label{eq:linearized-EL}
\end{equation}
The exact response operator is therefore
\begin{equation}
\chi := \frac{\delta q}{\delta J},
\label{eq:chi-def}
\end{equation}
and, whenever the local inverse exists,
\begin{equation}
\boxed{
\chi
=
-\,\left(
\frac{\delta^2 \Gamma[\bar q]}{\delta q\,\delta q}
\right)^{-1}.
}
\label{eq:chi-inverse-hessian}
\end{equation}
All inverse response operators in what follows are understood on the responsive tangent quotient, or equivalently on the range after gauge, conservation-law, and other null directions have been removed. Equivalently, on the source-side one has
\begin{equation}
\chi
=
\frac{\delta^2 E[J]}{\delta J\,\delta J},
\qquad
\frac{\delta^2 \Gamma[\bar q]}{\delta q\,\delta q}
=
-\,\chi^{-1},
\label{eq:W-Gamma-hessian}
\end{equation}
with the usual sign conventions understood.

This identity is the exact two-point content of the effective theory. The point is not merely that the Hessian can be inverted to produce a response function. The point is that the inverse response is itself a variational object: up to sign, it is the Hessian of the effective functional associated with the chosen basic variable. Once this is recognized, the natural question is no longer simply what diagrams or matrix equations are called ``RPA,'' but rather what part of this exact stiffness kernel is retained and what part is discarded in an RPA closure.

It is equally important to note what this statement does not imply. The Hessian controls exact linear response, but it does not exhaust the full effective theory. Higher derivatives,
\begin{equation}
\frac{\delta^3 \Gamma}{\delta q\,\delta q\,\delta q},
\qquad
\frac{\delta^4 \Gamma}{\delta q\,\delta q\,\delta q\,\delta q},
\qquad
\dots,
\label{eq:higher-vertices}
\end{equation}
control nonlinear response and higher point structure. Thus the Hessian is not the whole many-body problem. It is the exact two-point part of that problem, and this is precisely the level at which RPA must be defined.\cite{Baym1962,CJT1974}

\subsection{Exact Hessian decomposition}

Exact response is still not RPA. To define an RPA approximation, one must first specify how the exact Hessian is decomposed.

A sufficient realization of the Hessian decomposition is provided when the effective functional admits a decomposition of the form
\begin{equation}
\Gamma[q]
=
\Gamma_{\mathrm{ref}}[q]
+
\frac{1}{2}\,\pair{q}{K_{\mathrm{int}}q}
+
\Phi[q].
\label{eq:Gamma-decomposition}
\end{equation}
Here \(\Gamma_{\mathrm{ref}}\) is a chosen reference functional, \(K_{\mathrm{int}}\) is an explicitly retained bilinear interaction kernel in the chosen variable, and \(\Phi[q]\) is the irreducible remainder. At the exact level this is not yet an approximation. It is simply a decomposition of the theory into a reference part, an explicit bilinear interaction, and everything else. For the RPA closure itself, however, this stronger global functional decomposition is not required. It is sufficient that, at the chosen background, the exact Hessian admit the corresponding channel-adapted local decomposition; the retained kernel need not integrate to a unique global functional contribution.

In physically meaningful realizations, neither part of the retained Hessian is chosen arbitrarily. The reference functional is normally taken from a stable reference theory, often the noninteracting one, so that its Hessian defines a well-posed reference inverse response. The interaction kernel \(K_{\mathrm{int}}\) is then taken to be the linearized mean-field interaction kernel appropriate to the response channel of interest. In standard realizations, this retained kernel is often much simpler than the full two-point correction.

The remainder \(\Phi[q]\) is not assumed to begin at cubic order in fluctuations. Rather, it is defined only relative to the chosen decomposition, and may contribute already at the two-point level through its background-dependent Hessian
\begin{equation}
\frac{\delta^2 \Phi[\bar q]}{\delta q\,\delta q}.
\label{eq:Phi-background-hessian}
\end{equation}
Thus the exact linear response theory is naturally parameterized by the chosen background configuration \(\bar q\). In this sense, the practical content of an RPA closure is that one keeps the specified reference-plus-mean-field part of the Hessian, namely the reference contribution together with \(K_{\mathrm{int}}\), while the remaining irreducible contribution is collected into \(\Phi[q]\). Whether this is accurate is a further physical question, but the structural idea is already clear at this stage.

Differentiating twice at the physical configuration \(\bar q\) gives
\begin{equation}
\frac{\delta^2 \Gamma[\bar q]}{\delta q\,\delta q}
=
\frac{\delta^2 \Gamma_{\mathrm{ref}}[\bar q]}{\delta q\,\delta q}
+
K_{\mathrm{int}}
+
\frac{\delta^2 \Phi[\bar q]}{\delta q\,\delta q}.
\label{eq:Hessian-decomposition}
\end{equation}
This equation is the key structural input for what follows. The first term is the reference stiffness. The second term is the explicitly retained interaction. The third term is the exact irreducible remainder at the two-point level. It is this last term that carries the part of the exact Hessian not already contained in the retained reference plus interaction structure.

At this point the phrase ``what RPA neglects'' acquires a precise meaning. RPA is not the omission of response itself, nor of the Hessian as such. It is the omission of the irreducible remainder in the exact Hessian decomposition. Without this decomposition, the phrase ``RPA kernel'' remains too vague to identify a definite approximation.

\subsection{Abstract definition of RPA}

\begin{definition}[RPA as a Hessian closure]
Consider an effective functional \(\Gamma[q]\) with exact Hessian decomposition
\begin{equation}
\frac{\delta^2 \Gamma[\bar q]}{\delta q\,\delta q}
=
\frac{\delta^2 \Gamma_{\mathrm{ref}}[\bar q]}{\delta q\,\delta q}
+
K_{\mathrm{int}}
+
\frac{\delta^2 \Phi[\bar q]}{\delta q\,\delta q}.
\label{eq:exact-H-decomp}
\end{equation}
RPA is the closure obtained by discarding the irreducible remainder at the Hessian level:
\begin{equation}
\boxed{
\frac{\delta^2 \Gamma_{\mathrm{RPA}}[\bar q]}{\delta q\,\delta q}
:=
\frac{\delta^2 \Gamma_{\mathrm{ref}}[\bar q]}{\delta q\,\delta q}
+
K_{\mathrm{int}}.
}
\label{eq:RPA-Hessian}
\end{equation}
The corresponding response is
\begin{equation}
\boxed{
\chi_{\mathrm{RPA}}
=
-\,\left(
\frac{\delta^2 \Gamma_{\mathrm{RPA}}[\bar q]}{\delta q\,\delta q}
\right)^{-1}.
}
\label{eq:RPA-response}
\end{equation}
\end{definition}

Here RPA does not denote an arbitrary truncation of the exact Hessian: for a specified basic variable, background, reference theory, and response channel, \(K_{\mathrm{int}}\) is the corresponding linearized mean-field interaction kernel, whereas \(\delta^2\Phi[\bar q]/\delta q\,\delta q\) is the remaining irreducible Hessian contribution omitted by the closure.

\section{Four variational realizations of RPA: static density, dynamical density, equal-time bilocal, and spacetime-bilocal levels}
\label{sec:levels}

We now realize the abstract Hessian framework of Section~\ref{sec:framework} in four concrete settings:
static DFT, LR-TDDFT, 1RDMFT, and MBPT.  These correspond respectively to
the static local density, the dynamical density, the equal-time bilocal
1RDM, and the full spacetime-bilocal Green's function.  We begin with the
static density level, where the structure is most transparent.





\subsection{Density level: the Lieb functional, static response, and RPA screening}
\label{subsec:dftlevel}

At the density level, the natural source is a local scalar potential and the natural basic variable is the density. The corresponding dual pair may be written abstractly as \((\Xrho,\Xrho^\ast)\), with density \(\rho\in \Xrho\), potential \(v\in \Xrho^\ast\), and pairing
\begin{equation}
\pair{v}{\rho}
=
\int \dd\mathbf r\,v(\mathbf r)\,\rho(\mathbf r).
\label{eq:dft_pairing}
\end{equation}
For Coulombic systems, a standard choice is recalled in Appendix~\ref{app:domains}.

The universal density functional is
\begin{equation}
F[\rho]
=
\inf_{\Gamma\mapsto \rho}
\Tr\!\bigl[\Gamma(\hat T+\hat W)\bigr],
\label{eq:F_lieb}
\end{equation}
defined on the appropriate admissible density domain.\cite{Levy1979,Lieb1983}
In the static fixed-particle-number formulation used below, the trial density matrices in this constrained search are understood to be supported on the chosen \(N\)-particle sector \(\mathcal H_N\); the subscript \(N\) on \(F_N\) is suppressed for notational simplicity.
Given an external local potential \(v(\mathbf r)\), the source-side energy functional is
\begin{equation}
E[v]
=
\inf_\rho
\left\{
F[\rho]
+
\int \dd\mathbf r\,
v(\mathbf r)\,\rho(\mathbf r)
\right\},
\label{eq:E_density}
\end{equation}
and the corresponding variable-side functional is
\begin{equation}
\Gamma_{\mathrm{DFT}}[\rho]=F[\rho]
=
\sup_v
\left\{
E[v]
-
\int \dd\mathbf r\,v(\mathbf r)\,\rho(\mathbf r)
\right\}.
\label{eq:Gamma_density}
\end{equation}

If one considers a physical problem in a fixed-particle-number sector, the Euler equation takes the form
\begin{equation}
\frac{\delta F[\rho]}{\delta \rho(\mathbf r)}
=
-\,v(\mathbf r)+\mu.
\label{eq:Euler_density_F}
\end{equation}
Linearizing within that sector gives
\begin{equation}
\int \dd\mathbf r'\,
\frac{\delta^2 F[\rho]}
{\delta \rho(\mathbf r)\,\delta \rho(\mathbf r')}
\,\delta \rho(\mathbf r')
=
-\,\delta v(\mathbf r)+\delta\mu.
\label{eq:linearized_density_lieb}
\end{equation}
Equivalently, one may restrict density variations to the fixed-particle-number tangent space
\(
\{\delta\rho:\int \delta\rho=0\}
\)
and identify potentials modulo additive constants. In that quotient-space formulation, the constant \(\delta\mu\) is absorbed into the representative of \(\delta v\). All fixed-\(N\) inverse-response relations below are understood in this sense.
Defining the exact static density response by
\begin{equation}
\delta \rho = \chi\,\delta v,
\label{eq:chi_density}
\end{equation}
one obtains
\begin{equation}
\frac{\delta^2 F[\rho]}{\delta \rho\,\delta \rho}
=
-\,\chi^{-1}.
\label{eq:F_hessian}
\end{equation}
Equivalently,
\begin{equation}
\frac{\delta^2 E[v]}
{\delta v\,\delta v}
=
\chi.
\label{eq:E_hessian}
\end{equation}

In kernel notation, this response is the static density-density response function
\begin{equation}
\chi(\mathbf r,\mathbf r')
=
\frac{\delta \rho(\mathbf r)}
{\delta v(\mathbf r')}.
\label{eq:static_density_response_kernel}
\end{equation}
Thus the Hessian of \(F[\rho]\) is, up to sign, the inverse static density response.

To identify the RPA closure, decompose \(F[\rho]\) in the standard density-functional form
\begin{equation}
F[\rho]
=
T_{\mathrm{s}}[\rho]
+
E_{\mathrm{H}}[\rho]
+
E_{\mathrm{xc}}[\rho],
\label{eq:F_decomposition}
\end{equation}
where
\begin{equation}
E_{\mathrm{H}}[\rho]
=
\frac12
\iint \dd\mathbf r\,\dd\mathbf r'\,
\frac{\rho(\mathbf r)\rho(\mathbf r')}{|\mathbf r-\mathbf r'|}.
\label{eq:Hartree_density_lieb}
\end{equation}
Differentiating twice yields
\begin{equation}
\frac{\delta^2 F[\rho]}{\delta \rho\,\delta \rho}
=
\frac{\delta^2 T_{\mathrm{s}}[\rho]}{\delta \rho\,\delta \rho}
+
\frac{\delta^2 E_{\mathrm{H}}[\rho]}{\delta \rho\,\delta \rho}
+
\frac{\delta^2 E_{\mathrm{xc}}[\rho]}{\delta \rho\,\delta \rho}.
\label{eq:hessian_decomp_density_lieb}
\end{equation}
The Hartree term gives the Coulomb kernel
\begin{equation}
\frac{\delta^2 E_{\mathrm{H}}[\rho]}{\delta \rho(\mathbf r)\,\delta \rho(\mathbf r')}
=
\frac{1}{|\mathbf r-\mathbf r'|}
=:v(\mathbf r,\mathbf r'),
\label{eq:Coulomb_kernel_lieb}
\end{equation}
while the exchange-correlation kernel is
\begin{equation}
f_{\mathrm{xc}}(\mathbf r,\mathbf r')
:=
\frac{\delta^2 E_{\mathrm{xc}}[\rho]}
{\delta \rho(\mathbf r)\,\delta \rho(\mathbf r')}.
\label{eq:fxc_lieb}
\end{equation}
Since \(T_{\mathrm{s}}[\rho]\) is the noninteracting reference functional, the noninteracting static response \(\chi_0\) satisfies
\begin{equation}
\frac{\delta^2 T_{\mathrm{s}}[\rho]}{\delta \rho\,\delta \rho}
=
-\,\chi_0^{-1},
\label{eq:Ts_hessian_lieb}
\end{equation}
and therefore
\begin{equation}
\boxed{
\chi^{-1}
=
\chi_0^{-1}
-
v
-
f_{\mathrm{xc}}.
}
\label{eq:exact_density_hessian_lieb}
\end{equation}

This is the static density-level realization of the general Hessian decomposition of Section~\ref{sec:framework}. Here
\begin{equation}
\Gamma_{\mathrm{ref}}[\rho]=T_{\mathrm{s}}[\rho],
\qquad
K_{\mathrm{int}}=v,
\qquad
\Phi[\rho]=E_{\mathrm{xc}}[\rho].
\label{eq:density_identification_lieb}
\end{equation}
Direct density-response RPA is obtained by discarding the exchange-correlation remainder:
\begin{equation}
\boxed{
f_{\mathrm{xc}}\approx 0.
}
\label{eq:density_rpa_closure_lieb}
\end{equation}
Equivalently,
\begin{equation}
\boxed{
\chi_{\mathrm{RPA}}
=
(\chi_0^{-1}-v)^{-1}.
}
\label{eq:density_rpa_response_lieb}
\end{equation}
At the static density level, this closure gives the zero-frequency density
response and the corresponding static RPA screening, including the familiar
static Lindhard/RPA dielectric response of the electron gas.\cite{Lindhard1954,FetterWalecka1971,GiulianiVignale2005}

To connect the static density Hessian construction with the adiabatic
connection, introduce the fixed-density coupling-constant path
\begin{align}
F^\lambda[\rho]
&=
\inf_{\Gamma\mapsto\rho}
\Tr\!\bigl[\Gamma(\hat T+\lambda\hat W)\bigr],
\qquad
\lambda\in[0,1],
\label{eq:Flambda_lieb}
\\
F^0[\rho]&=T_{\mathrm s}[\rho],
\qquad
F^1[\rho]=F[\rho].
\label{eq:Flambda_endpoints}
\end{align}
When differentiability along this path is available, the exchange-correlation
energy is obtained from the density adiabatic connection,
\begin{equation}
E_{\mathrm{xc}}[\rho]
=
\int_0^1
\left(
\Tr\!\bigl[\Gamma^\lambda_\rho\hat W\bigr]
-
E_{\mathrm H}[\rho]
\right)d\lambda .
\label{eq:AC_density_lieb}
\end{equation}
Using the fluctuation-dissipation theorem, this adiabatic-connection
expression can be rewritten directly in terms of the imaginary-frequency
density response as
\begin{equation}
E_{\mathrm{xc}}^{\mathrm{ACFD}}[\rho]
=
-\frac{1}{2\pi}
\int_0^1 d\lambda
\Tr\!\left[
v \bigl( \int_0^\infty d\omega\, \chi^\lambda(i\omega)+\pi \rho\delta\bigr)
\right],
\label{eq:ACFD_main}
\end{equation}
with the usual convention for the density-response trace and the Coulomb
self term. Equivalently, the correlation part may be written as
\begin{equation}
E_{\mathrm c}^{\mathrm{ACFD}}[\rho]
=
-\frac{1}{2\pi}
\int_0^1 d\lambda
\int_0^\infty d\omega\,
\Tr\!\left[
v\bigl(\chi^\lambda(i\omega)-\chi_0(i\omega)\bigr)
\right].
\label{eq:ACFD_correlation_main}
\end{equation}
The derivation is recorded in Appendix~\ref{app:FDT}. In the direct RPA
closure, the response entering Eq.~\eqref{eq:ACFD_correlation_main} is
approximated by
\begin{equation}
\chi^\lambda_{\mathrm{RPA}}(i\omega)
=
\left[
\chi_0(i\omega)^{-1}
-
\lambda v
\right]^{-1}.
\label{eq:ACFD_RPA_chi}
\end{equation}
This motivates the dynamical density level discussed next.

\subsection{Dynamical density level: LR-TDDFT and density-channel RPA}
\label{subsec:lrtdftlevel}

The dynamical density level keeps the density as the basic variable, but
lifts the scalar source from a static local potential to an imaginary-time-dependent local source. Thus the enlargement from DFT to LR-TDDFT is not a
spatial bilocalization of the source. It is a dynamical enrichment of the
density channel. The source is a one-time field and may break imaginary-time
translation symmetry away from equilibrium; the response is evaluated at an
equilibrium background, where simultaneous imaginary-time translation
invariance is restored. The corresponding dynamical density dual pair
\((\Xrhobeta,\Xrhobetastar)\) and its pairing are recorded in
Appendix~\ref{app:td_density_domain}. The grand-canonical meaning of
\(\beta\) and \(\mu\) is summarized in
Appendix~\ref{app:grandcanonical_parameters}.

We use a Euclidean source-functional formulation because it gives a clean
Hessian and Legendre-transform structure. The corresponding real-time
LR-TDDFT formulation can also be written in an action or Keldysh-contour
language, where causality of the response is manifest; for the present
Hessian discussion, however, the Matsubara formulation is the most economical
one.\cite{RungeGross1984,vanLeeuwen1998,Vignale2008}

Let
\[
\hat K=\hat H-\mu\hat N,
\]
and let \(j\in\Xrhobetastar\) be a local imaginary-time density source. With
the sign convention chosen so that the source derivative gives the density,
define
\begin{align}
Z_{\beta,\mu}[j]
&=
\TrF\Biggl(
e^{-\beta\hat K}
T_\tau
\exp\Bigl[
-\int_0^\beta d\tau
\int \dd\mathbf r\,
j(\mathbf r,\tau)\hat\rho_{\mathrm H}(\mathbf r,\tau)
\Bigr]
\Biggr),
\label{eq:Z_beta_j_td}
\end{align}
where
\[
\hat\rho_{\mathrm H}(\mathbf r,\tau)
=
e^{\tau\hat K}\hat\rho(\mathbf r)e^{-\tau\hat K}.
\]
Set
\begin{equation}
E_{\beta,\mu}[j]
:=
-\ln Z_{\beta,\mu}[j].
\label{eq:E_beta_j_td}
\end{equation}
Then
\begin{equation}
\rho
=
\frac{\delta E_{\beta,\mu}[j]}{\delta j},
\label{eq:rho_from_j_td}
\end{equation}
and the corresponding dynamical density effective action is
\begin{equation}
\Gamma_{\beta,\mu}[\rho]
=
\sup_j
\left\{
E_{\beta,\mu}[j]
-
\pair{j}{\rho}
\right\}.
\label{eq:Gamma_beta_rho_td}
\end{equation}

Linearizing the stationarity condition gives the exact Matsubara density
response
\begin{equation}
\chi_{\beta,\mu}
:=
\frac{\delta \rho}{\delta j},
\label{eq:chi_beta_td_def}
\end{equation}
and hence
\begin{equation}
\frac{\delta^2\Gamma_{\beta,\mu}[\rho]}
{\delta\rho\,\delta\rho}
=
-\chi_{\beta,\mu}^{-1}.
\label{eq:Gamma_beta_hessian_td}
\end{equation}
This is the dynamical density-channel analogue of
Eq.~\eqref{eq:F_hessian}. The inverse is understood on the responsive quotient after conserved or gauge-like null directions have been removed. The source is local in spacetime, but the response kernel is a two-point spacetime object.

At an equilibrium background, simultaneous imaginary-time translation
invariance implies
\(\chi(\mathbf r\tau,\mathbf r'\tau')=
\chi(\mathbf r,\mathbf r';\tau-\tau')\). The response may therefore be
represented in bosonic Matsubara frequency. The exact inverse-response
decomposition has the same Hessian form as in the static density theory:
\begin{equation}
\boxed{
\chi(i\omega)^{-1}
=
\chi_0(i\omega)^{-1}
-
v
-
f_{\mathrm{xc}}(i\omega).
}
\label{eq:td_exact_inverse_response}
\end{equation}
Here \(\chi_0(i\omega)\) is the noninteracting dynamical density response,
\(v\) is the instantaneous Coulomb kernel, and
\(f_{\mathrm{xc}}(i\omega)\) is the dynamical exchange-correlation Hessian.
The distinction from the static density level is precisely that the exact
exchange-correlation kernel may be frequency dependent, reflecting memory
effects.

This formulation also separates two common approximations. In an adiabatic
TDDFT approximation, one replaces the dynamical exchange-correlation kernel
by an instantaneous, frequency-independent kernel obtained from a
ground-state functional.\cite{Petersilka1996,Maitra2016} Direct
density-channel RPA is stronger: it discards the exchange-correlation
Hessian altogether,
\begin{equation}
\boxed{
f_{\mathrm{xc}}(i\omega)\approx 0.
}
\label{eq:td_rpa_closure}
\end{equation}
Equivalently,
\begin{equation}
\boxed{
\chi_{\mathrm{RPA}}(i\omega)
=
\left[
\chi_0(i\omega)^{-1}
-
v
\right]^{-1}.
}
\label{eq:td_rpa_response}
\end{equation}

Along the density adiabatic-connection path used in
Section~\ref{subsec:dftlevel}, the same formulas are read with
\(v\) replaced by \(\lambda v\) and with the exact response and kernel
replaced by \(\chi^\lambda(i\omega)\) and
\(f^\lambda_{\mathrm{xc}}(i\omega)\). Inserting
\(\chi^\lambda_{\mathrm{RPA}}(i\omega)\) into
Eq.~\eqref{eq:ACFD_correlation_main} gives the usual ACFD RPA correlation energy.

\subsection{Equal-time bilocal level: 1RDMFT and static nonlocal Hessian closures}
\label{subsec:1rdmlevel}

The equal-time bilocal branch enlarges the basic variable from the local density to the one-body reduced density matrix. This is the spatially nonlocal enrichment of the density-level theory, in contrast to the dynamical enrichment described in Section~\ref{subsec:lrtdftlevel}. The corresponding dual pair may be written abstractly as \((\Xgamma,\Xgamma^\ast)\), with \(\gamma\in \Xgamma\), \(u\in \Xgamma^\ast\), and pairing
\begin{equation}
\pair{u}{\gamma}
=
\iint \dd \mathbf r\,\dd \mathbf r'\,
u(\mathbf r',\mathbf r)\,
\gamma(\mathbf r,\mathbf r').
\label{eq:gamma_u_pairing}
\end{equation}
A natural finite-kinetic-energy trace-class realization of this pair is recorded in Appendix~\ref{app:domains}.
The conjugate source at this level is a Hermitian equal-time bilocal one-body potential
\begin{equation}
u(\mathbf r,\mathbf r'),
\qquad
u(\mathbf r,\mathbf r')=u^*(\mathbf r',\mathbf r).
\label{eq:u_def}
\end{equation}
A local scalar potential is recovered by the special choice
\begin{equation}
u(\mathbf r',\mathbf r)=v(\mathbf r)\,\delta(\mathbf r-\mathbf r'),
\label{eq:u_local_restriction}
\end{equation}
for which
\begin{equation}
\iint \dd \mathbf r\,\dd \mathbf r'\,
u(\mathbf r',\mathbf r)\,
\gamma(\mathbf r,\mathbf r')
=
\int \dd \mathbf r\,v(\mathbf r)\,\gamma(\mathbf r,\mathbf r)
=
\int \dd \mathbf r\,v(\mathbf r)\,\rho(\mathbf r).
\label{eq:u_local_to_density_pairing}
\end{equation}

Introduce the one-body reduced density matrix
\begin{equation}
\gamma(\mathbf r,\mathbf r')
=
\Tr\!\left[
\Gamma\,
\hat\psi^\dagger(\mathbf r')\hat\psi(\mathbf r)
\right],
\label{eq:gamma_def}
\end{equation}
and the corresponding universal functional
\begin{equation}
F[\gamma]
=
\inf_{\Gamma\mapsto\gamma}
\Tr\!\left[
\Gamma(\hat T+\hat W)
\right],
\label{eq:F_gamma}
\end{equation}
defined on the appropriate admissible 1RDM domain.\cite{Coleman1963,Gilbert1975,Valone1980,Blochl2013}
As at the density level, this static functional is understood as a constrained search over density matrices supported on \(\mathcal H_N\), with the particle-number subscript suppressed.

The source-side energy functional is
\begin{equation}
E[u]
=
\inf_\gamma
\left\{
F[\gamma]
+
\iint \dd \mathbf r\,\dd \mathbf r'\,
u(\mathbf r',\mathbf r)\,
\gamma(\mathbf r,\mathbf r')
\right\},
\label{eq:Eu_gamma}
\end{equation}
and the corresponding variable-side functional is
\begin{equation}
\Gamma_{\mathrm{1RDM}}[\gamma]=F[\gamma]
=
\sup_u
\left\{
E[u]
-
\iint \dd \mathbf r\,\dd \mathbf r'\,
u(\mathbf r',\mathbf r)\,
\gamma(\mathbf r,\mathbf r')
\right\}.
\label{eq:Gamma_1RDM}
\end{equation}

If one considers a fixed-particle-number physical problem, the Euler equation becomes
\begin{equation}
\frac{\delta F[\gamma]}
{\delta \gamma(\mathbf r,\mathbf r')}
=
-\,u(\mathbf r',\mathbf r)
+
\mu\,\delta(\mathbf r-\mathbf r').
\label{eq:Euler_gamma}
\end{equation}
Linearizing within that sector gives
\begin{equation}
\iint \dd \mathbf s\,\dd \mathbf s'\,
\frac{\delta^2 F[\gamma]}
{\delta \gamma(\mathbf r,\mathbf r')\,
 \delta \gamma(\mathbf s,\mathbf s')}
\,\delta \gamma(\mathbf s,\mathbf s')
=
-\,\delta u(\mathbf r',\mathbf r)
+\delta\mu\,\delta(\mathbf r-\mathbf r').
\label{eq:linearized_gamma}
\end{equation}
Equivalently, one may work on the trace-zero tangent space
\(\{\delta\gamma:\Trf\delta\gamma=0\}\)
and identify one-body sources modulo multiples of the identity. The \(\delta\mu\) term is then absorbed into the source representative, and the inverse below is understood on the corresponding quotient space.
At zero temperature the ensemble-representable domain is constrained by \(0\leq\gamma\leq1\), and physically important noninteracting states often lie on its boundary. The ordinary Hessian identity below is therefore asserted only on a smooth, locally parametrized stratum, or in a finite-dimensional regularization where the source-to-1RDM map is locally invertible. At occupation-boundary or idempotent points, the appropriate object is a constrained Hessian or subdifferential geometry. Appendix~\ref{app:grandcanonical_DFT_1RDMFT} records the finite-temperature formulation. At finite temperature and finite orbital dimension, entropy keeps the occupations strictly between zero and one and supplies a nontrivial reference curvature.

On such a smooth branch, define the exact equal-time bilocal response by
\begin{equation}
\Lambda(\mathbf r,\mathbf r';\mathbf s,\mathbf s')
:=
\frac{\delta \gamma(\mathbf r,\mathbf r')}
{\delta u(\mathbf s',\mathbf s)}.
\label{eq:Lambda_exact}
\end{equation}
Then
\begin{equation}
\frac{\delta^2 F[\gamma]}{\delta \gamma\,\delta \gamma}
=
-\,\Lambda^{-1}.
\label{eq:gamma_Hessian_response}
\end{equation}
At finite temperature, the parallel identity is written for the intrinsic free-energy functional \(A_{\beta}[\gamma]\). Its noninteracting reference part includes the entropy curvature, so \(A_{\mathrm{s},\beta}''=-\Lambda_{\mathrm{s},\beta}^{-1}\) need not vanish. The chemical potential enters the external linear term \(-\mu\Trf\gamma\), rather than the intrinsic functional itself. The zero-temperature limit may become singular as occupations approach zero or one.

To identify the exact Hessian decomposition, write
\begin{equation}
F[\gamma]
=
T[\gamma]
+
E_{\mathrm{H}}[\rho]
+
E_{\mathrm{xc}}[\gamma],
\label{eq:F_gamma_decomp_compact}
\end{equation}
where
\begin{equation}
T[\gamma]
=
\left.
-\frac{1}{2}
\int \dd \mathbf r\,
\nabla_{\mathbf r'}^2\,
\gamma(\mathbf r,\mathbf r')
\right|_{\mathbf r'=\mathbf r},
\label{eq:T_gamma}
\end{equation}
\begin{equation}
\rho(\mathbf r)=\gamma(\mathbf r,\mathbf r),
\label{eq:rho_from_gamma}
\end{equation}
\begin{equation}
E_{\mathrm{H}}[\rho]
=
\frac{1}{2}
\iint \dd \mathbf r\,\dd \mathbf r'\,
\frac{\rho(\mathbf r)\rho(\mathbf r')}{|\mathbf r-\mathbf r'|},
\label{eq:EH_gamma}
\end{equation}
and
\begin{equation}
E_{\mathrm{xc}}[\gamma]
:=
W[\gamma]-E_{\mathrm{H}}[\rho],
\label{eq:Exc_gamma_def}
\end{equation}
with
\begin{equation}
W[\gamma]
=
\inf_{\Gamma\mapsto\gamma}
\Tr\!\left[
\Gamma\hat W
\right].
\label{eq:W_gamma}
\end{equation}
A fixed-\(\gamma\) coupling-constant path can also be introduced in analogy with the density adiabatic connection, but it is not needed for the static equal-time Hessian closure considered here.\cite{GiesbertzUimonenvanLeeuwen2018,Senjean2022}

It is useful to make the density contraction explicit. Let
\(D:X_\gamma\to X_\rho\) denote the diagonal map
\[
(D\gamma)(r)=\gamma(r,r)=\rho(r).
\]
Then the Hartree term may be written as
\[
E_{\mathrm{H}}[\rho]
=
E_{\mathrm{H}}[D\gamma]
=
\frac12\langle D\gamma,vD\gamma\rangle ,
\]
so that
\[
\frac{\delta^2E_{\mathrm{H}}[D\gamma]}
{\delta\gamma\,\delta\gamma}
=
D^\ast vD .
\]

At the level of the ambient zero-temperature functional, the kinetic term is affine in \(\gamma\), and its ordinary Hessian vanishes:
\[
\frac{\delta^2T[\gamma]}{\delta\gamma\,\delta\gamma}=0.
\]
This ambient identity isolates the interaction kernels retained below. It should not be read as a complete formula for the noninteracting 1RDM inverse response at idempotent or occupation-boundary points, where orbital rotations, energy denominators, and the curvature of the representability constraints enter through constrained variational geometry. Thus the canonical 1RDMFT decomposition
\[
F[\gamma]
=
T[\gamma]+E_{\mathrm{H}}[D\gamma]+E_{\mathrm{xc}}[\gamma]
\]
gives
\[
\frac{\delta^2F[\gamma]}{\delta\gamma\,\delta\gamma}
=
D^\ast vD
+
\frac{\delta^2E_{\mathrm{xc}}[\gamma]}
{\delta\gamma\,\delta\gamma}.
\]

The 1RDM-level RPA closure is obtained by discarding the exchange-correlation Hessian,
\[
\frac{\delta^2E_{\mathrm{xc}}[\gamma]}
{\delta\gamma\,\delta\gamma}
\approx 0.
\]
Hence
\[
\frac{\delta^2F_{\mathrm{RPA}}[\gamma]}
{\delta\gamma\,\delta\gamma}
=
D^\ast vD .
\]
Since \(D^\ast vD\) acts only through the diagonal density contraction, the direct Hartree-retained closure defines only a stiffness on the density-coupled quotient \(X_\gamma/\ker D\). It leaves genuinely off-diagonal 1RDM directions undetermined and does not by itself define a complete bilocal response operator. A complete response requires either an additional retained bilocal kernel, as below, or the appropriate finite-temperature or constrained reference curvature.

One may also define an exchange-inclusive closure by splitting
\[
E_{\mathrm{xc}}[\gamma]
=
E_{\mathrm{x}}[\gamma]+E_{\mathrm{c}}[\gamma].
\]
For the Hartree--Fock exchange functional,
\[
E_{\mathrm{x}}[\gamma]
=
-\frac12
\iint dx\,dx'\,
\frac{\gamma(x,x')\gamma(x',x)}{|r-r'|},
\]
where \(x=(r,\sigma)\), the exchange Hessian is the bilocal kernel
\[
K_{\mathrm{x}}
:=
\frac{\delta^2E_{\mathrm{x}}[\gamma]}
{\delta\gamma\,\delta\gamma}.
\]
Equivalently, in the present index convention,
\[
(K_{\mathrm{x}}\eta)(x,x')
=
-\frac{\eta(x',x)}{|r-r'|}.
\]

The exchange-inclusive closure, denoted here by xRPA, is obtained by discarding only the correlation Hessian,
\[
\frac{\delta^2E_{\mathrm{c}}[\gamma]}
{\delta\gamma\,\delta\gamma}
\approx 0.
\]
The retained Hessian is then
\[
\frac{\delta^2F_{\mathrm{xRPA}}[\gamma]}
{\delta\gamma\,\delta\gamma}
=
D^\ast vD+K_{\mathrm{x}} .
\]
Unlike the direct Hartree kernel, \(K_{\mathrm{x}}\) acts on bilocal 1RDM variations. Thus the exchange-inclusive closure is not restricted to the density-coupled channel. When the retained Hessian is invertible in the chosen 1RDM response space, the corresponding response is
\[
\Lambda_{\mathrm{xRPA}}
=
-\left(D^\ast vD+K_{\mathrm{x}}\right)^{-1}.
\]
This inverse is a formal branch-wise response in the chosen ambient or regularized 1RDM space. At idempotent or occupation-boundary points, it must be replaced by the corresponding constrained Hessian or by a finite-temperature construction with nontrivial reference curvature.

At the 1RDM level, the terms RPA and xRPA are proposed here only in the structural sense introduced in Section~\ref{sec:framework}; they are not standard names for a complete zero-temperature 1RDM response theory. RPA denotes the direct Hartree-retained Hessian closure, while xRPA denotes the exchange-inclusive closure obtained by retaining the Hartree--Fock exchange Hessian as well.

\subsection{Spacetime-bilocal level: MBPT and two-particle Hessian closures}
\label{subsec:mbptlevel}

At the full spacetime-bilocal level, the natural basic variable is the one-particle Green's function. This level combines the two enrichments discussed above: it is bilocal in space, as in the 1RDM level, and dynamical in time, as in the LR-TDDFT level. We use the compact notation
\(1=(\mathbf r_1,\tau_1,\sigma_1)\), \(2=(\mathbf r_2,\tau_2,\sigma_2)\),
with \(d1\) denoting integration over space and imaginary time and summation over spin.

Let \(J(2,1)\) be a bilocal c-number generating source coupled to the one-body operator \(\hat\psi^\dagger(2)\hat\psi(1)\). To distinguish the physical domain from its linearized geometry, we regard the admissible Green's functions as lying in an affine domain modeled locally on a linear variation space \(\XG\). Bilocal source variations belong to the continuous dual \(\XG^\ast\), with pairing
\begin{equation}
\langle J,G\rangle
=
\int d1\,d2\,J(2,1)G(1,2).
\label{eq:JG_pairing_MBPT}
\end{equation}
Thus, if the complete Green's function \(G(1,2)\) is to be obtained as a first derivative of a generating functional, the conjugate c-number source must retain the same bilocal index structure.

The physical equilibrium background is taken at auxiliary source \(J=0\), where the Green's function is invariant under simultaneous imaginary-time translations,
\begin{equation}
\bar G(x_1\tau_1,x_2\tau_2)
=
\bar G(x_1,x_2;\tau_1-\tau_2),
\label{eq:G_TTI_background}
\end{equation}
and is diagonal in fermionic Matsubara frequency. The source variations used to generate the full two-particle response are nevertheless allowed to be general bilocal variations near this background and may break time-translation invariance. This is required to probe finite bosonic transfer frequencies; restricting the variations themselves to the time-translation-invariant sector would retain only the zero-transfer sector of the Hessian. In particular, the auxiliary source is not assumed to be time local: \(\tau_1\) and \(\tau_2\) are independent arguments, and the full bilocal source is required if \(G(1,2)\) itself is to be the first-derivative variable.

Time-local Hermitian kernels form only the physical instantaneous one-body-perturbation subspace of this larger auxiliary source space. Such a source has the form
\begin{equation}
J_u(2,1)
=
\delta(\tau_2-\tau_1)\,u(x_2,x_1;\tau_1),
\qquad
u(x_2,x_1;\tau)=u^\ast(x_1,x_2;\tau),
\label{eq:J_timelocal_MBPT}
\end{equation}
with \(u\) Hermitian, and its time-independent spatially bilocal and spatially local time-dependent subclasses recover the static 1RDMFT and LR-TDDFT sources, respectively. At an equilibrium Hermitian background, the standard thermal-covariance argument makes the source Hessian negative semidefinite along these physical directions. This is a restricted Hessian statement, not a proof of concavity for arbitrary bilocal source directions. Bilocality itself is not the obstruction: fully bilocal symmetric sources may retain concavity in positive-measure Euclidean field theories. The distinction in the fermionic construction is that the Berezin functional integral is not a positive probability measure. The local branch construction used below is therefore stated separately from this restricted covariance statement.

The corresponding domain conventions are summarized in Appendix~\ref{app:domains}. The key structural requirement is that \(\XG\) admit a continuous equal-time reduction to \(\Xgamma\), and that \(\XG^\ast\) contain the time-local embeddings of \(\Xgamma^\ast\). The grand-canonical meaning of \(\beta\) and \(\mu\) is summarized in Appendix~\ref{app:grandcanonical_parameters}.

Set
\[
\hat K=\hat H-\mu\hat N .
\]
The source-dependent partition functional may be written as
\begin{align}
Z_\beta[J]
&=
\TrF\Biggl(
e^{-\beta\hat K}
\,T_\tau
\exp\Bigl[
-\int d1\,d2\,
J(2,1)\,
\hat\psi_\mathrm{H}^\dagger(2)\hat\psi_\mathrm{H}(1)
\Bigr]
\Biggr),
\label{eq:ZJ_MBPT_operator}
\end{align}
where the Heisenberg fields are defined with respect to \(\hat K\). In the present paper, source derivatives, Legendre transforms, determinant formulas, and diagrammatic constructions at the MBPT level are defined first for finite-dimensional Berezin regularizations obtained by time slicing and one-particle truncation. Continuum notation is used only as shorthand for the corresponding limiting structure.

The natural source-side functional at this level is
\begin{equation}
E_\beta[J] := -\ln Z_\beta[J].
\label{eq:E_MBPT}
\end{equation}
At the equilibrium background, the standard Duhamel--Kubo--Mori covariance argument makes the Hessian of \(E_\beta[J]\) negative semidefinite along time-local Hermitian one-body source directions; the corresponding local statement is recorded in Appendix~\ref{app:MBPTsource}. It does not establish concavity on the full spacetime-bilocal auxiliary source space. The issue is not bilocality alone. In positive-measure Euclidean field theories, the Hessian is a negative variance and may be sign definite even for fully bilocal symmetric sources. In the fermionic Berezin formulation, by contrast, no positive probability measure underlies the source integral, and the general bilocal Hessian need not have a definite sign.\cite{LinLindsey2018}

The generic fully bilocal construction therefore starts from local Legendre geometry rather than a global extremal principle. Here \(\XG^\ast\) denotes a chosen real form of admissible bilocal kernels, compatible with the Euclidean reality conditions and with the selected real branch of the source functional; its complexification may instead be treated by complex-analytic local geometry. We select a differentiable source branch \(\mathcal B\subset\XG^\ast\) containing the equilibrium point \(J=0\), and assume only that the source-to-Green's-function map is locally invertible on the responsive subspace. We choose the sign of the bilocal source in Eq.~\eqref{eq:ZJ_MBPT_operator} so that
\begin{equation}
G(1,2)
=
\frac{\delta E_\beta[J]}{\delta J(2,1)},
\label{eq:G_from_W}
\end{equation}
with \(G(1,2)=-\langle T_\tau\hat\psi_\mathrm{H}(1)\hat\psi_\mathrm{H}^\dagger(2)\rangle_J\). The transpose in the source argument is part of the dual pairing \(\int J(2,1)G(1,2)\). Let \(J=J[G]\) denote the local inverse on the image \(\mathcal G_{\mathcal B}:=\nabla E_\beta(\mathcal B)\). The branch-wise effective functional is
\begin{align}
\Gamma_{\mathcal B}[G]
&=
E_\beta[J[G]]
-
\int d1\,d2\,
J[G](2,1)\,G(1,2).
\label{eq:GammaG_local}
\end{align}
It obeys \(\delta\Gamma_{\mathcal B}/\delta G=-J[G]\) and \(\Gamma_{\mathcal B}''=-(E_\beta'')^{-1}\) wherever the local inverse exists. No convexity or global extremal characterization is assumed for the complete bilocal fermionic source space.

If, on a restricted convex source domain \(\mathcal D\subset\mathcal B\), the functional \(E_\beta\) is additionally proper and concave and the stationary source is the global maximizer, then the same local effective functional admits the stronger Legendre--Fenchel representation
\begin{equation}
\Gamma[G]
=
\sup_{J\in\mathcal D}
\left\{
E_\beta[J]-\langle J,G\rangle
\right\}.
\label{eq:GammaG_LF_conditional}
\end{equation}
Restricted physical Hermitian source sectors and positive-measure Euclidean theories illustrate how such additional positivity can arise; the full Legendre--Fenchel representation requires the stronger domain and attainment conditions stated above. In the remainder of this section, the branch label is suppressed.
The physical equilibrium problem corresponds to the auxiliary source \(J=0\), and the physical Green's function \(\bar G\) is therefore determined by the stationarity condition
\begin{equation}
\frac{\delta \Gamma[\bar G]}{\delta G}=0.
\label{eq:stationarity_G}
\end{equation}

It is convenient to rewrite this abstract effective functional in the standard 2PI form. In the present setting, this form is obtained by isolating the explicit quadratic reference contribution in the finite-dimensional Legendre transform and collecting the remainder into the Luttinger--Ward interaction functional \(\Phi[G]\). A concise derivation is recorded in Appendix~\ref{app:MBPTsource}. One then writes
\begin{align}
\Gamma[G]
&=
-\,\Tr\!\bigl[(G_0^{-1}-G^{-1})G\bigr]
+
\Tr\ln(-G)
+
\Phi[G]
+
\text{const.}
\label{eq:GammaG_explicit}
\end{align}
The reference part is therefore
\begin{align}
\Gamma_{\mathrm{ref}}[G]
&=
-\,\Tr\!\bigl[(G_0^{-1}-G^{-1})G\bigr]
+
\Tr\ln(-G),
\label{eq:GammaG_ref}
\end{align}
where \(G_0\) is the noninteracting propagator.\cite{LuttingerWard1960,Klein1961,CJT1974,Almbladh1999}

Differentiating Eq.~\eqref{eq:GammaG_explicit} gives
\begin{equation}
\frac{\delta \Gamma[G]}{\delta G}
=
G^{-1}-G_0^{-1}
+
\frac{\delta \Phi[G]}{\delta G}.
\label{eq:dGamma_dG_clean}
\end{equation}
Defining the self-energy by
\begin{equation}
\Sigma[G]
:=
\frac{\delta \Phi[G]}{\delta G},
\label{eq:Sigma_from_Phi_clean}
\end{equation}
the stationarity condition becomes Dyson's equation
\begin{equation}
G^{-1}
=
G_0^{-1}
-
\Sigma[G].
\label{eq:Dyson_MBPT_clean}
\end{equation}

The exact linear response object at this level is the generalized two-particle response
\begin{equation}
L(12;34)
:=
\frac{\delta G(1,2)}{\delta J(4,3)}.
\label{eq:L_exact}
\end{equation}
At the equilibrium background, \(L\) is invariant under a simultaneous shift of all four imaginary times. In frequency space it is therefore parameterized by two fermionic frequencies and one conserved bosonic transfer frequency. The nonzero-transfer sectors arise from source variations that break time-translation invariance around the time-translation-invariant background.
Hence
\begin{equation}
\frac{\delta^2 \Gamma[G]}{\delta G\,\delta G}
=
-\,L^{-1}.
\label{eq:GammaG_hessian}
\end{equation}
Define the exact irreducible two-particle kernel by
\begin{equation}
K_{\mathrm{irr}}(12;34)
:=
\frac{\delta^2 \Phi[G]}
{\delta G(1,2)\,\delta G(3,4)}.
\label{eq:Kirr_MBPT}
\end{equation}
Then the exact Hessian decomposition at the Green's function level is
\begin{equation}
\boxed{
\frac{\delta^2 \Gamma[G]}{\delta G\,\delta G}
=
\frac{\delta^2 \Gamma_{\mathrm{ref}}[G]}{\delta G\,\delta G}
+
K_{\mathrm{irr}}.
}
\label{eq:GammaG_hessian_exact_MBPT}
\end{equation}

To exhibit the direct RPA-type closure, one chooses the retained interaction kernel in the direct particle-hole channel to be the bare Coulomb kernel \(v\), and writes
\begin{equation}
K_{\mathrm{irr}} = v + K_{\mathrm{rem}}.
\label{eq:Kirr_split_MBPT}
\end{equation}
This is a chosen Hessian-level decomposition adapted to the direct channel. It is not a claim that the full Luttinger--Ward functional admits a unique channel-independent splitting into a bare part and a remainder.\cite{Potthoff2006,KozikFerreroGeorges2015,Vucicevic2018} Exchange-inclusive variants replace \(v\) by the corresponding antisymmetrized Hartree--Fock kernel, whereas closures that retain additional irreducible correlation kernels are more appropriately regarded as RPA-based or beyond-RPA extensions.

Accordingly,
\begin{equation}
\boxed{
\frac{\delta^2 \Gamma[G]}{\delta G\,\delta G}
=
\frac{\delta^2 \Gamma_{\mathrm{ref}}[G]}{\delta G\,\delta G}
+
v
+
K_{\mathrm{rem}}.
}
\label{eq:GammaG_hessian_split}
\end{equation}
This is the Green's function-level realization of the general scheme. Here the retained interaction kernel is
\begin{equation}
K_{\mathrm{int}}=v,
\label{eq:Kint_MBPT}
\end{equation}
while the interaction contribution to the exact Hessian enters through
\begin{equation}
\frac{\delta^2 \Phi[G]}{\delta G\,\delta G}
=
K_{\mathrm{irr}}
=
v+K_{\mathrm{rem}}.
\label{eq:mbpt_identification}
\end{equation}

Direct RPA is obtained by discarding the irreducible remainder:
\begin{equation}
\boxed{
K_{\mathrm{rem}} \approx 0.
}
\label{eq:MBPT_RPA_closure}
\end{equation}
Thus
\begin{equation}
\boxed{
\frac{\delta^2 \Gamma_{\mathrm{RPA}}[G]}{\delta G\,\delta G}
=
\frac{\delta^2 \Gamma_{\mathrm{ref}}[G]}{\delta G\,\delta G}
+
v.
}
\label{eq:GammaG_RPA_hessian}
\end{equation}
The associated RPA response is
\begin{equation}
\boxed{
L_{\mathrm{RPA}}
=
-\,\left(
\frac{\delta^2 \Gamma_{\mathrm{RPA}}[G]}{\delta G\,\delta G}
\right)^{-1}.
}
\label{eq:L_RPA_hessian}
\end{equation}

All Green's-function Hessians in this subsection are evaluated at a specified background \(\bar G\). If one defines
\begin{equation}
L_0
:=
-\,\left(
\frac{\delta^2 \Gamma_{\mathrm{ref}}[\bar G]}{\delta G\,\delta G}
\right)^{-1},
\label{eq:L0_def}
\end{equation}
then \(L_0\) is the independent-particle two-particle propagator constructed from that background. A prescribed noninteracting, Kohn--Sham, or Hartree--Fock \(\bar G\) gives the corresponding reference RPA, whereas a dressed \(\bar G\) leads to a self-consistent variant. The exact and approximate Hessian relations may then be rewritten in the equivalent response-equation forms
\begin{align}
L
&=
L_0
+
L_0\,(v+K_{\mathrm{rem}})\,L,
\label{eq:BSE_L}
\\
L_{\mathrm{RPA}}
&=
L_0
+
L_0\,v\,L_{\mathrm{RPA}}.
\label{eq:L_RPA}
\end{align}
These Dyson or Bethe--Salpeter forms are used here only as response-equation representations of the underlying Hessian closure.

A coupling-constant interpolation of the Luttinger--Ward functional may also
be introduced by replacing \(\hat W\) by \(\lambda\hat W\), giving a family
\(\Phi^\lambda[G]\). This additional structure is not needed for the present
Hessian closure. At the Green's function level, the RPA content relevant here is
the retained two-particle kernel and the resulting response equation.

\subsection{Projection, reduction, and the four-corner hierarchy}
\label{subsec:reduction}

The four variational levels are connected by exact forward reductions of
variables and corresponding restrictions of sources. Schematically,
\begin{equation}
\begin{array}{ccc}
& G(\mathbf r\tau,\mathbf r'\tau') & \\[0.3em]
\swarrow & & \searrow \\[0.3em]
\gamma(\mathbf r,\mathbf r') & & \rho(\mathbf r,\tau) \\[0.3em]
\searrow & & \swarrow \\[0.3em]
& \rho(\mathbf r) &
\end{array}
\label{eq:four_corner_variable_diagram}
\end{equation}
where the left branch removes dynamical information first and then takes the
diagonal density, while the right branch first projects to the local
time-dependent density channel and then removes the time dependence.

The source restrictions are the dual operations. A full spacetime-bilocal
source \(J(2,1)\) reduces on the static bilocal branch to
\(J_u(2,1)=u(x_2,x_1)\delta(\tau_2-\tau_1)\), which is the 1RDMFT source, or on the dynamical density branch to
\(J_j(2,1)=j(x_1,\tau_1)\delta(x_2-x_1)\delta(\tau_2-\tau_1)\), which is the LR-TDDFT source. The latter may break time-translation invariance away from equilibrium. The static DFT source is obtained by further restricting \(j\) to be time independent.

At the response level, these reductions are linear contractions. With the
notation of Appendix~\ref{app:MBPTsource},
\begin{equation}
\Lambda=\Pet\,L\,\Iet,
\qquad
\chi=\Pdiag\,\Pet\,L\,\Idiag.
\label{eq:response_reductions_four}
\end{equation}
Here \(\Pet\) and \(\Pdiag\) act on variables, while \(\Iet\) and
\(\Idiag\) embed the corresponding restricted sources into the full
spacetime-bilocal source space. The first contraction is understood in the static, zero-transfer sector appropriate to 1RDMFT, whereas the second retains arbitrary bosonic transfer frequency. The static density response is obtained from the zero-frequency, or time-integrated, sector of this dynamical density response.

This restriction can strengthen the variational geometry. In general, let \(R:X\to Y\) be a linear reduction and \(R^\ast:Y^\ast\to X^\ast\) its dual source embedding. The reduced source functional is
\begin{equation}
E_Y[j]=E_X[R^\ast j],
\qquad
E_Y''[j]=R\,E_X''[R^\ast j]R^\ast.
\label{eq:source_restriction_hessian}
\end{equation}
An indefinite Hessian on the full source space may therefore become sign definite on a distinguished reduced source channel, although this is not automatic. The relevant strengthening is supplied by additional positivity of the reduced physical sector. For example, a fully bilocal fermionic Green's-function theory may possess only local saddle-type Legendre geometry, whereas finite-temperature equal-time 1RDM and density channels admit genuine concave source functionals and Legendre--Fenchel duals. Their zero-temperature limits may retain convexity while losing differentiability at occupation or representability boundaries.

On the variable side, reduction also eliminates unresolved directions. Under the standard lower-semicontinuity, closure, and constraint-qualification assumptions, a globally convex reduction is represented by the infimal projection,
\begin{equation}
\Gamma_Y[y]
=
\inf_{x:\,Rx=y}\Gamma_X[x],
\label{eq:infimal_reduction_geometry}
\end{equation}
while in a generic local saddle setting the same operation is understood as stationary elimination on a chosen branch.

The corresponding relation at the Hessian level is subtler, because the
Hessian is the inverse response. Reducing a response by projection is not the
same as projecting its inverse. To make this explicit, introduce local
coordinates \(Q=(q,h)\), where \(q\) is retained and \(h\) is eliminated, and
write the full Hessian as
\begin{equation}
H
=
\begin{pmatrix}
H_{qq} & H_{qh}\\
H_{hq} & H_{hh}
\end{pmatrix}.
\label{eq:block_hessian_reduction}
\end{equation}
If the hidden stationary branch \(h=h[q]\) is locally well defined and
\(H_{hh}\) is invertible on the eliminated subspace, then the Hessian of the
reduced effective functional is
\begin{equation}
\boxed{
H_{\mathrm{red}}
=
H_{qq}-H_{qh}H_{hh}^{-1}H_{hq}.
}
\label{eq:schur_hessian_reduction}
\end{equation}
Thus the reduction map on inverse response is a Schur complement, which is
nonlinear in the blocks of \(H\). In particular, for a retained projection
\(\mathcal P\), one generally has
\begin{equation}
\mathcal S\!\left(H_{\mathrm{ref}}+K_{\mathrm{int}}\right)
\neq
\mathcal S\!\left(H_{\mathrm{ref}}\right)
+
\mathcal P K_{\mathrm{int}}\mathcal P^\ast,
\label{eq:closure_reduction_noncommute}
\end{equation}
where \(\mathcal S\) denotes elimination by Schur complement. This gives the
local mathematical reason that Hessian closure and exact reduction need not
commute.

This is why RPA closures at different corners of the hierarchy need not be
identical. DFT RPA, LR-TDDFT RPA, 1RDMFT-level Hessian closures, and MBPT RPA
are parallel realizations of the same Hessian principle after different
choices of basic variable and response channel, not the same approximation
written in four notations.

\section{Discussion and conclusion}
\label{sec:discussion}

The main conclusion of this work is that RPA is most naturally understood as
a Hessian closure. Once a source and a basic variable have been chosen, the
exact effective functional determines the exact inverse response through its
Hessian. RPA is obtained by retaining a reference Hessian together with an
explicit interaction kernel and discarding the remaining irreducible
contribution.

This viewpoint separates three ingredients that are often compressed in
standard presentations: the choice of variable, the exact response theory
associated with that variable, and the closure that defines the approximation.
At the static density level, the closure gives the familiar inverse-response
decomposition underlying zero-frequency density response and static RPA
screening. At the dynamical density level, it gives LR-TDDFT RPA, where the
exchange-correlation kernel may carry frequency dependence and where the RPA
response enters the ACFD correlation-energy formula. At the equal-time
bilocal level, the direct Hartree-retained closure acts only through the
diagonal density channel, while exchange-inclusive closures retain additional
bilocal structure. At the spacetime-bilocal level, the same logic appears as
a closure of the two-particle Hessian of the Green's function effective
functional.

The four levels are connected by exact forward reductions of variables and
corresponding restrictions of sources, but their RPA closures are not
identical. Projection is simple for response kernels, whereas inverse-response
Hessians reduce through effective elimination of hidden directions. Thus DFT
RPA, LR-TDDFT RPA, 1RDMFT-level Hessian closures, and MBPT RPA should be
viewed as parallel realizations of one Hessian principle, not as the same
approximation written in different notation.

The hierarchy is also accompanied by a hierarchy of variational strength. The common structure is the local source--variable duality and inverse-Hessian relation. A generic fully bilocal fermionic theory may have an indefinite Hessian and only a branch-wise stationary Legendre formulation. Restriction to physical source channels and elimination of hidden variables can expose additional positivity, yielding locally convex or concave reduced theories. With further global regularity and attainment, these become genuine Legendre--Fenchel variational principles. Convexity is therefore an enhanced realization of the Hessian framework, not a universal prerequisite for it.

This interpretation also clarifies the position of related approximations.
Exchange-inclusive RPA, generalized RPA, and QRPA correspond to different
choices of variable, channel, and retained kernel in the exact Hessian
decomposition. \(GW\) is not itself a definition of RPA, but a downstream
one-particle self-energy approximation built from a screened interaction
whose polarization sector is often obtained from an RPA-type closure.
Likewise, \(G_0W_0\) is best viewed as a one-shot quasiparticle correction
built on an RPA-screened interaction, not as another form of the RPA Hessian
closure itself.\cite{Hedin1965,Ren2012}

A complete convex-analytic treatment of nondifferentiability,
representability restrictions, and singular Hessians, together with a
numerical assessment of less standard 1RDM- and Green's-function-level
closures, lies beyond the structural scope of the present work. These issues
do not change the main message: RPA is not fundamentally a diagrammatic
slogan, a density-functional trick, or a matrix equation tied to one
formalism. We advocate a unifying characterization of a broad class of RPA constructions as closures of the exact Hessian of an effective functional.

\appendix 

\section{Admissible domains and dual pairings}
\label{app:domains}

This appendix records the domain conventions actually used in the main text.
The goal is not to give a complete functional-analytic construction at each
level, but to make explicit the dual spaces, pairings, and admissible sets
underlying the source--variable dualities employed in the paper. The
dynamical density level is treated as an imaginary-time lift of the static
density source--variable pairing.

\subsection{Density level}

At the density level, one works with a dual pair \((\Xrho,\Xrho^\ast)\) of density and potential spaces. For Coulombic systems in three dimensions, a standard choice is
\begin{equation}
\Xrho = L^1(\R^3)\cap L^3(\R^3),
\qquad
\Xrho^\ast = L^{3/2}(\R^3)+L^\infty(\R^3),
\label{eq:Xrho_app}
\end{equation}
with pairing
\begin{equation}
\pair{v}{\rho}
=
\int \dd\mathbf r\,v(\mathbf r)\rho(\mathbf r).
\label{eq:dft_pairing_app}
\end{equation}
The variable-side functional is then the Lieb functional, understood as a convex lower semicontinuous functional on \(\Xrho\), while the source-side object is the concave ground-state energy functional on \(\Xrho^\ast\).\cite{Lieb1983} Fixed particle-number sectors are recovered by imposing \(\int \rho = N\) in the physical variational problem.

\subsection{Dynamical density level}
\label{app:td_density_domain}

At the dynamical density level, the static density source--variable pairing is lifted to imaginary time. A convenient working convention is
\begin{equation}
\Xrhobeta
\sim
L^p_{\mathrm{per}}([0,\beta];\Xrho),
\qquad
\Xrhobetastar
\sim
L^{p'}_{\mathrm{per}}([0,\beta];\Xrho^\ast),
\label{eq:Xrho_beta_app}
\end{equation}
with \(p^{-1}+{p'}^{-1}=1\). The subscript ``per'' indicates the bosonic imaginary-time periodicity appropriate for density sources and density responses. The pairing is
\begin{equation}
\pair{j}{\rho}
=
\int_0^\beta d\tau
\int \dd\mathbf r\,
j(\mathbf r,\tau)\rho(\mathbf r,\tau).
\label{eq:td_density_pairing_app}
\end{equation}
Static DFT is recovered by restricting to the time-independent subspace.

This notation is meant as a working Banach-space convention rather than a complete functional-analytic construction of time-dependent density-representability. The structural requirements used in the main text are only that \(\Xrhobeta\) contain admissible imaginary-time density variations, that \(\Xrhobetastar\) contain admissible time-dependent scalar sources, and that the pairing in Eq.~\eqref{eq:td_density_pairing_app} be finite and well defined.

\subsection{Equal-time bilocal level}

At the equal-time bilocal level, a convenient admissible set is the ensemble-representable set of self-adjoint one-body density matrices satisfying
\begin{equation}
\gamma=\gamma^\dagger,
\qquad
0\le \gamma \le 1,
\label{eq:gamma_constraints_app_1}
\end{equation}
together with a finite-kinetic-energy condition such as
\begin{equation}
\Tr\!\bigl[(1+\hat T)\gamma\bigr]<\infty.
\label{eq:finite_kinetic_app}
\end{equation}
If one wishes to restrict to a fixed particle-number sector, one further imposes
\begin{equation}
\Tr \gamma = N.
\label{eq:gamma_constraints_app_2}
\end{equation}
For the present paper, one may take \(\Xgamma\) to be the real linear span of such admissible positive operators. The dual space \(\Xgamma^\ast\) is correspondingly understood as a suitable class of Hermitian one-body potentials or Hermitian sesquilinear forms bounded relative to the kinetic-energy graph norm.\cite{Coleman1963,Valone1980,Gilbert1975}

The natural pairing is
\begin{equation}
\pair{u}{\gamma}
=
\Trf(u\gamma)
=
\iint \dd \mathbf r\,\dd \mathbf r'\,
u(\mathbf r',\mathbf r)\,
\gamma(\mathbf r,\mathbf r').
\label{eq:gamma_pairing_app}
\end{equation}
Ordinary local scalar potentials are recovered as the diagonal subclass
\begin{equation}
u(\mathbf r',\mathbf r)=v(\mathbf r)\delta(\mathbf r-\mathbf r').
\label{eq:local_subclass_app}
\end{equation}

\subsection{Spacetime-bilocal level}

At the Green's function level, the corresponding functional-analytic structure
is more delicate. The physical time-ordered fermionic Green's function is not
most naturally viewed as an arbitrary regular bilocal kernel, because it has
a fixed equal-time discontinuity. With the convention
\(G(1,2)=-\langle T_\tau\hat\psi(1)\hat\psi^\dagger(2)\rangle\), one has
\begin{equation}
G(x,\tau^+;x',\tau)-G(x,\tau^-;x',\tau)
=
-\,\delta(x-x'),
\label{eq:jump_app}
\end{equation}
where \(x=(\mathbf r,\sigma)\). At a time-translation-invariant equilibrium background, the Green's function depends only on the relative imaginary time and is diagonal in fermionic Matsubara frequency. In that representation, physical Green's functions share the universal high-frequency asymptotics
\begin{equation}
G(i\omega_n)
=
\frac{1}{i\omega_n}I
+
O(\omega_n^{-2}).
\label{eq:high_freq_tail_app}
\end{equation}
The tail tends to zero as \(|\omega_n|\to\infty\), but it is the frequency
representation of the equal-time jump.

It is therefore convenient to treat the Green's function domain as an affine
space. We write
\begin{equation}
G = G_{\mathrm{can}}+\widetilde G,
\label{eq:G_split_app}
\end{equation}
where \(G_{\mathrm{can}}\) is a fixed canonical part carrying the universal
equal-time discontinuity, for example with leading behavior
\begin{equation}
G_{\mathrm{can}}(i\omega_n)=\frac{1}{i\omega_n}I,
\label{eq:Gcan_app}
\end{equation}
and where \(\widetilde G\) belongs to a linear space of admissible regular
remainders. This notation is only a domain convention; it is not meant to
construct a complete Lieb-type Green's function functional space. The source--Green's-function pairing is defined first in the finite-dimensional regularization, where both the canonical affine part and the regular remainder are included; the continuum notation denotes the corresponding regularized limit whenever it exists.

Motivated by the 1RDM space \(\Xgamma\), one may regard the regular part of
the Green's function variation space schematically as a time lift of
\(\Xgamma\):
\begin{equation}
\XG \sim \XT \,\widehat\otimes\, \Xgamma,
\qquad
\XG^\ast \sim \XT^\ast \,\widehat\otimes\, \Xgamma^\ast,
\label{eq:XG_candidate_app}
\end{equation}
where \(\XT\) denotes an appropriate space of bilocal kernels on \([0,\beta]^2\), with the fermionic boundary conditions appropriate to the two time arguments. The only structural requirements needed in the
present paper are the following:
\begin{enumerate}[label=(\roman*), leftmargin=2.5em]
\item the physical Green's function is treated as an affine object
\(G_{\mathrm{can}}+\XG\);
\item there is a well-defined equal-time reduction map
\(\Pet:\XG\to\Xgamma\);
\item the diagonal density reduction \(\Pdiag:\Xgamma\to\Xrho\) is well
defined;
\item the source space \(\XG^\ast\) contains the time-local embeddings of
\(\Xgamma^\ast\).
\end{enumerate}
The equilibrium background lies in the time-translation-invariant subspace, but the variation spaces \(\XG\) and \(\XG^\ast\) must also contain directions that break time-translation invariance in order to describe nonzero bosonic transfer frequencies. In particular, if \(u\in \Xgamma^\ast\) is a static equal-time one-body source, then the spacetime-bilocal source
\begin{equation}
J(2,1)=u(x_2,x_1)\,\delta(\tau_2-\tau_1)
\label{eq:time_local_source_app}
\end{equation}
should define an element of \(\XG^\ast\).

The precise construction of a canonical Banach pair \((\XG,\XG^\ast)\) is
left open here and is part of the broader prospective Lieb-style program for
Green's function theory mentioned in Section~\ref{sec:discussion}.\cite{Lieb1983,LinLindsey2018}

\section{From dynamical density response to the ACFD RPA energy}
\label{app:FDT}

This appendix records how the density adiabatic-connection integrand is
rewritten in terms of the imaginary-frequency density response. The main
text uses only the final ACFD expression.

\subsection{Pair fluctuation form of the adiabatic-connection integrand}

For a fixed coupling constant \(\lambda\), let
\(\Gamma^\lambda_\rho\) be a minimizing state at fixed density \(\rho\). Define
\begin{equation}
\delta\hat\rho(\mathbf r)
=
\hat\rho(\mathbf r)-\rho(\mathbf r)
\end{equation}
and the equal-time density-fluctuation correlation function
\begin{equation}
C^\lambda(\mathbf r,\mathbf r')
=
\left\langle
\delta\hat\rho(\mathbf r)\delta\hat\rho(\mathbf r')
\right\rangle_\lambda .
\label{eq:C_lambda_app}
\end{equation}
Using
\begin{equation}
\hat W
=
\frac12
\iint d\mathbf r\,d\mathbf r'\,
v(\mathbf r,\mathbf r')
\left[
\hat\rho(\mathbf r)\hat\rho(\mathbf r')
-
\delta(\mathbf r-\mathbf r')\hat\rho(\mathbf r)
\right],
\label{eq:W_density_app}
\end{equation}
one obtains
\begin{equation}
\Tr\!\bigl[\Gamma^\lambda_\rho\hat W\bigr]
-
E_{\mathrm H}[\rho]
=
\frac12
\Tr\!\left[
v\bigl(C^\lambda-\rho\delta\bigr)
\right].
\label{eq:AC_integrand_C_app}
\end{equation}
Therefore the exchange-correlation energy may be written as
\begin{equation}
E_{\mathrm{xc}}[\rho]
=
\frac12
\int_0^1 d\lambda\,
\Tr\!\left[
v\bigl(C^\lambda-\rho\delta\bigr)
\right].
\label{eq:Exc_C_app}
\end{equation}

\subsection{Fluctuation-dissipation relation}

For a nondegenerate ground state, the imaginary-frequency density response has
the Lehmann representation~\cite{FetterWalecka1971,GiulianiVignale2005}
\begin{equation}
\chi^\lambda(\mathbf r,\mathbf r';i\omega)
=
-2
\sum_{m>0}
\frac{
\Omega_m^\lambda
\langle 0^\lambda|\delta\hat\rho(\mathbf r)|m^\lambda\rangle
\langle m^\lambda|\delta\hat\rho(\mathbf r')|0^\lambda\rangle
}{
(\Omega_m^\lambda)^2+\omega^2
},
\label{eq:chi_iw_Lehmann_app}
\end{equation}
where \(\Omega_m^\lambda=E_m^\lambda-E_0^\lambda\). Since
\begin{equation}
\int_0^\infty
\frac{d\omega}{\pi}
\frac{2\Omega}{\Omega^2+\omega^2}
=
1,
\end{equation}
the equal-time correlation function satisfies
\begin{equation}
C^\lambda(\mathbf r,\mathbf r')
=
-\int_0^\infty
\frac{d\omega}{\pi}\,
\chi^\lambda(\mathbf r,\mathbf r';i\omega).
\label{eq:FDT_zeroT_app}
\end{equation}

Substituting Eq.~\eqref{eq:FDT_zeroT_app} into
Eq.~\eqref{eq:Exc_C_app} gives
\begin{equation}
E_{\mathrm{xc}}^{\mathrm{ACFD}}[\rho]
=
-\frac{1}{2\pi}
\int_0^1 d\lambda
\Tr\!\left[
v \bigl( \int_0^\infty d\omega\, \chi^\lambda(i\omega)+\pi \rho\delta\bigr)
\right].
\label{eq:ACFD_app}
\end{equation}
The direct RPA version is obtained by replacing
\(\chi^\lambda(i\omega)\) with
\(\chi^\lambda_{\mathrm{RPA}}(i\omega)\). After subtracting the
\(\lambda=0\) contribution, this gives the common correlation-only form
\begin{equation}
E_{\mathrm c}^{\mathrm{RPA}}[\rho]
=
-\frac{1}{2\pi}
\int_0^1 d\lambda
\int_0^\infty d\omega\,
\Tr\!\left[
v\bigl(
\chi^\lambda_{\mathrm{RPA}}(i\omega)-\chi_0(i\omega)
\bigr)
\right].
\label{eq:Ec_RPA_app}
\end{equation}

\section{Grand-canonical ensemble parameters in imaginary-time source formulations}
\label{app:grandcanonical_parameters}

This appendix records the common grand-canonical convention used by the
imaginary-time source formulations in Sections~\ref{subsec:lrtdftlevel}
and~\ref{subsec:mbptlevel}. The parameters \(\beta\) and \(\mu\) have their
standard meaning as Lagrange multipliers for the average energy and average
particle number.

They arise from the maximum-entropy problem for density matrices on
fermionic Fock space. One maximizes the von Neumann entropy
\begin{equation}
S[\Gamma]
=
-\,\TrF\!\bigl(\Gamma\ln\Gamma\bigr)
\label{eq:entropy_app}
\end{equation}
subject to
\begin{equation}
\TrF \Gamma = 1,
\qquad
\TrF (\Gamma \hat H)=\bar E,
\qquad
\TrF (\Gamma \hat N)=\bar N.
\label{eq:constraints_app}
\end{equation}
The stationary solution is
\begin{equation}
\Gamma_{\beta,\mu}
=
Z_{\beta,\mu}^{-1}
e^{-\beta(\hat H-\mu \hat N)},
\label{eq:grand_canonical_state_app}
\end{equation}
with
\begin{equation}
Z_{\beta,\mu}
=
\TrF e^{-\beta(\hat H-\mu \hat N)}.
\label{eq:grand_partition_app}
\end{equation}
Thus \(\beta\) and \(\mu\) are the Lagrange multipliers associated with the average-energy and average-particle-number constraints.

The corresponding state equations are
\begin{equation}
\frac{\partial \ln Z_{\beta,\mu}}{\partial \mu}
=
\beta \,\langle \hat N\rangle_{\beta,\mu},
\qquad
\frac{\partial \ln Z_{\beta,\mu}}{\partial \beta}
=
-\,\langle \hat H-\mu \hat N\rangle_{\beta,\mu},
\label{eq:state_eqs_grand_app}
\end{equation}
hence
\begin{equation}
\langle \hat N\rangle_{\beta,\mu}
=
\frac{1}{\beta}
\frac{\partial \ln Z_{\beta,\mu}}{\partial \mu},
\qquad
\langle \hat H\rangle_{\beta,\mu}
=
-\,\frac{\partial \ln Z_{\beta,\mu}}{\partial \beta}
+
\mu\langle \hat N\rangle_{\beta,\mu}.
\label{eq:NH_from_logZ_grand_app}
\end{equation}
In the zero-temperature limit, the grand-canonical state projects onto the ground sector of \(\hat H-\mu \hat N\). To recover the \(N\)-particle ground state of \(\hat H\), one chooses \(\mu\) so that
\begin{equation}
E_0(N)-E_0(N-1)
\le \mu \le
E_0(N+1)-E_0(N).
\label{eq:mu_window_grand_app}
\end{equation}

This convention is common to the local density-source functional used in
LR-TDDFT and the bilocal Green's function source functional used in MBPT. The
additional finite-dimensional Berezin regularization needed for the
Green's function construction is recorded separately in
Appendix~\ref{app:MBPTsource}.

\section{MBPT source functionals, finite-dimensional Berezin regularization, and the 2PI effective functional}
\label{app:MBPTsource}

This appendix records the technical points underlying the MBPT discussion of
Section~\ref{subsec:mbptlevel}. The guiding choice is the following: source
derivatives, Legendre transforms, determinant formulas, and diagrammatic
constructions are defined first at the level of finite-dimensional Berezin
integrals obtained by time slicing and one-particle truncation. Continuum
notation is then used only as shorthand for the corresponding limiting
structure.

\subsection{Green's function source functional and finite-dimensional regularization}

Let
\begin{equation}
\hat K := \hat H-\mu \hat N,
\label{eq:Khat_def_app}
\end{equation}
and define the imaginary-time Heisenberg fields by
\begin{equation}
\hat\psi_\mathrm{H}(1)
=
e^{\tau_1 \hat K}\,
\hat\psi(\mathbf r_1,\sigma_1)\,
e^{-\tau_1 \hat K},
\qquad
\hat\psi_\mathrm{H}^\dagger(1)
=
e^{\tau_1 \hat K}\,
\hat\psi^\dagger(\mathbf r_1,\sigma_1)\,
e^{-\tau_1 \hat K},
\label{eq:Heisenberg_fields_app}
\end{equation}
with \(1\equiv(\mathbf r_1,\tau_1,\sigma_1)\). The source-dependent partition functional is
\begin{equation}
Z_\beta[J]
=
\TrF\Biggl(
e^{-\beta \hat K}
\,T_\tau
\exp\Bigl[
-\int d1\,d2\,
J(2,1)\,
\hat\psi_\mathrm{H}^\dagger(2)\hat\psi_\mathrm{H}(1)
\Bigr]
\Biggr).
\label{eq:Z_operator_app}
\end{equation}
At the physical equilibrium point \(J=0\), the Green's function is invariant under simultaneous imaginary-time translations and depends only on \(\tau_1-\tau_2\). The auxiliary source used in the 2PI construction is nevertheless a general bilocal kernel: no condition \(\tau_1=\tau_2\) is imposed, and unequal-time matrix elements are required to make the complete \(G(1,2)\) the first-derivative variable. A time-local source of the form
\(J_u(2,1)=u(x_2,x_1;\tau_1)\delta(\tau_2-\tau_1)\), with \(u(x_2,x_1;\tau)=u^*(x_1,x_2;\tau)\), is only the physical Hermitian one-body-perturbation subspace of this larger source space. Such a perturbation may depend on \(\tau\) and therefore break time-translation invariance away from equilibrium.

To regularize this expression, let \(P_N:\mathfrak h\to\mathfrak h\) be a rank-\(N\) projection on the one-particle Hilbert space, choose an orthonormal basis of \(P_N\mathfrak h\), and partition the imaginary-time interval \([0,\beta]\) into \(M\) slices. The resulting time-sliced coherent-state construction yields a finite-dimensional Berezin integral
\begin{equation}
Z_{\beta}^{(N,M)}[J]
=
\int
\prod_{\alpha=1}^{n}
d\bar\psi_\alpha\,d\psi_\alpha\,
\exp\!\left[-S_J^{(N,M)}(\bar\psi,\psi)\right],
\label{eq:ZNM_app}
\end{equation}
where \(n\) is the dimension of the discrete one-body spacetime index set. More explicitly, one may write the combined index as
\begin{equation}
\alpha=(a,m),
\label{eq:combined_spacetime_index_app}
\end{equation}
where \(a\) labels the retained spatial, orbital, spin, and other internal degrees of freedom and \(m\) labels the imaginary-time slice. Accordingly, \(G_{\alpha\beta}^{(N,M)}\) and \(J_{\beta\alpha}^{(N,M)}\) are matrices on the complete discretized one-particle spacetime index set, rather than merely orbital matrices. Physical time-local Hermitian perturbations form the structured subspace
\begin{equation}
J_{(b,m'),(a,m)}^{(N,M)}
=
\delta_{m'm}\,u_{ba}(m),
\qquad
u(m)=u(m)^\dagger,
\label{eq:J_timelocal_discrete_app}
\end{equation}
whereas the general matrix \(J^{(N,M)}\) is retained as the auxiliary source for the fully bilocal construction. The continuum notation used in the main text is shorthand for the corresponding limit
\begin{equation}
\int \mathcal D(\bar\psi,\psi)\,
e^{-S_J[\bar\psi,\psi]}
:=
\lim_{(N,M)\to(\infty,\infty)}
Z_{\beta}^{(N,M)}[J],
\label{eq:infinite_dim_def_app}
\end{equation}
whenever the relevant directed limit exists.

\subsection{Quadratic reference theory, source derivatives, and source-space concavity}

For the quadratic reference theory,
\begin{equation}
\hat H_0
=
\sum_{ab}
h_{ab}\,c_a^\dagger c_b,
\qquad
\hat K_0:=\hat H_0-\mu \hat N,
\label{eq:H0K0_app}
\end{equation}
the regularized quadratic action takes the form
\begin{equation}
S_0^{(N,M)}(\bar\psi,\psi)
=
\sum_{\alpha,\beta=1}^{n}
\bar\psi_\alpha
A_{0,\alpha\beta}^{(N,M)}
\psi_\beta,
\label{eq:S0_discrete_app}
\end{equation}
With the source convention used below, the regularized quadratic source action is
\begin{equation}
S_{0,J}^{(N,M)}(\bar\psi,\psi)
=
\bar\psi\bigl(A_0^{(N,M)}-J^{(N,M)}\bigr)\psi.
\label{eq:S0J_discrete_app}
\end{equation}
This sign convention is the one for which differentiation of \(E=-\ln Z\) gives \(G=\delta E/\delta J\). The free propagator is
\begin{equation}
G_0^{(N,M)}
:=
\bigl(A_0^{(N,M)}\bigr)^{-1}.
\label{eq:G0_discrete_app}
\end{equation}
In continuum notation one writes \(A_0=G_0^{-1}\).

With a bilocal c-number source \(J\), the quadratic reference partition function is written, in the determinant convention dual to Eq.~\eqref{eq:G_from_W}, as
\begin{equation}
Z_{0,\beta}^{(N,M)}[J]
=
\det\!\Bigl(
A_0^{(N,M)}-J^{(N,M)}
\Bigr),
\label{eq:Z0_det_app}
\end{equation}
hence
\begin{equation}
E_{0,\beta}^{(N,M)}[J]
:=
-\,\ln Z_{0,\beta}^{(N,M)}[J]
=
-\,\Tr\ln\!\Bigl(
A_0^{(N,M)}-J^{(N,M)}
\Bigr),
\label{eq:E0_trlog_app}
\end{equation}
up to the choice of logarithm branch.

Differentiation with respect to the c-number source is then an ordinary matrix derivative, and gives
\begin{equation}
G^{(N,M)}
=
\frac{\delta E_{0,\beta}^{(N,M)}[J]}
{\delta J^{(N,M)}}
=
\bigl(A_0^{(N,M)}-J^{(N,M)}\bigr)^{-1}.
\label{eq:G_from_E0_discrete_app}
\end{equation}
The same insertion formula holds for the interacting regularized theory:
\begin{equation}
\frac{\delta E_{\beta}^{(N,M)}[J]}
{\delta J_{\beta\alpha}^{(N,M)}}
=
G_{\alpha\beta}^{(N,M)}.
\label{eq:dE_dJ_discrete_app}
\end{equation}
At the equilibrium background \(J=0\), consider a time-local Hermitian direction
\begin{equation}
\delta J(2,1)
=
\delta(\tau_2-\tau_1)\,\delta u(x_2,x_1;\tau_1),
\qquad
\delta u(\tau)=\delta u(\tau)^\dagger.
\label{eq:physical_direction_concavity_app}
\end{equation}
Let \(\delta\hat O(\tau)\) denote the corresponding Hermitian one-body insertion. The second variation is the negative Duhamel--Kubo--Mori quadratic form, equivalently the integrated connected imaginary-time-ordered covariance,
\begin{align}
\left.
\frac{d^2}{dt^2}
E_{\beta}^{(N,M)}[t\,\delta J]
\right|_{t=0}
&=
-
\int_0^\beta d\tau\,d\tau'\,
\nonumber\\[-1mm]
&\quad\times
\bigl\langle
T_\tau\,\delta\hat O(\tau)\,\delta\hat O(\tau')
\bigr\rangle_{0}^{\mathrm{conn}}
\le 0.
\label{eq:concavity_discrete_app}
\end{align}
Thus the Hessian is negative semidefinite at equilibrium along this physical source subspace. This statement is local in source space; it is not a proof of global concavity even on the entire time-local source domain.

Bilocality by itself does not prevent a concavity theorem. In a finite-dimensional Euclidean field theory with an ordinary positive Gibbs measure, a symmetric bilocal perturbation produces a real quadratic random variable, and the corresponding Hessian is its negative variance. This is the setting in which rigorous global Green's-function convex duality can be established.\cite{LinLindsey2018} The fermionic coherent-state integral is different: Berezin integration is an algebraic Grassmann operation rather than integration against a positive probability measure. Already in the quadratic regularized theory, the identity \(\delta G_J=G_JH G_J\) gives, for a general bilocal matrix direction \(H\),
\begin{align}
D^2E_{0,\beta}^{(N,M)}[J](H,H)
&=
\Tr\!\left(G_JHG_JH\right),
\label{eq:general_bilocal_hessian_app}\\
G_J
&=
\left(A_0^{(N,M)}-J^{(N,M)}\right)^{-1}.
\nonumber
\end{align}
which is not generally of the positive form \(\Tr(X^\dagger X)\) and has no fixed sign on the full bilocal matrix space. The Matsubara Green's function obeys the reality relation \(G(i\omega_n)^\dagger=G(-i\omega_n)\), rather than being Hermitian and positive at each frequency; even imposing an adjoint condition on \(H\) does not by itself make \(G_JH\) Hermitian or positive. Consequently, the covariance argument above does not extend to arbitrary bilocal matrices coupling unequal imaginary-time slices.

For the fully bilocal fermionic construction, the generic conclusion is therefore local rather than globally convex. We work on a chosen real form of the regularized bilocal source space, compatible with the Euclidean reality conditions and with the selected real branch of \(E_\beta^{(N,M)}[J]\). We select a differentiable branch \(\mathcal B^{(N,M)}\) containing \(J=0\) and assume that the regularized map \(J\mapsto G[J]\) is locally invertible on the responsive subspace. No sign definiteness of the full bilocal Hessian is assumed.

\subsection{Local Legendre transform, reference functional, and the Luttinger--Ward remainder}

Let \(J=J[G]\) denote the local inverse on the image of \(\mathcal B^{(N,M)}\). The regularized branch-wise effective functional is
\begin{equation}
\Gamma_{\beta,\mathcal B}^{(N,M)}[G]
=
E_{\beta}^{(N,M)}[J[G]]
-
\Tr\!\bigl(J[G]G\bigr).
\label{eq:Gamma_discrete_app}
\end{equation}
It obeys the usual local dual relations, independently of convexity. If the source functional is additionally concave on a restricted convex domain \(\mathcal D^{(N,M)}\subset\mathcal B^{(N,M)}\) and the stationary source is the global maximizer, the same functional has the enhanced representation
\begin{equation}
\Gamma_{\beta}^{(N,M)}[G]
=
\sup_{J\in\mathcal D^{(N,M)}}
\left\{
E_{\beta}^{(N,M)}[J]-\Tr(JG)
\right\}.
\label{eq:Gamma_discrete_LF_app}
\end{equation}
We suppress the branch label below. For the quadratic reference theory one can carry out the local transform explicitly. Since
\begin{equation}
G^{(N,M)}
=
\bigl(A_0^{(N,M)}-J^{(N,M)}\bigr)^{-1},
\label{eq:GJ_relation_app}
\end{equation}
one has
\begin{equation}
J^{(N,M)}
=
A_0^{(N,M)}-(G^{(N,M)})^{-1}.
\label{eq:J_from_G_discrete_app}
\end{equation}
Substituting into the local Legendre transform gives the regularized reference functional
\begin{equation}
\Gamma_{\mathrm{ref},\beta}^{(N,M)}[G]
=
-\,\Tr\!\bigl[
\bigl(A_0^{(N,M)}-G^{-1}\bigr)G
\bigr]
+
\Tr\ln G
+
\text{const.}
\label{eq:Gamma_ref_discrete_app}
\end{equation}
or, equivalently in continuum notation,
\begin{equation}
\Gamma_{\mathrm{ref}}[G]
=
-\,\Tr\!\bigl[(G_0^{-1}-G^{-1})G\bigr]
+
\Tr\ln(-G)
+
\text{const.}
\label{eq:Gamma_ref_minusG_app}
\end{equation}

The interaction remainder is then defined by
\begin{equation}
\Phi[G]
:=
\Gamma[G]-\Gamma_{\mathrm{ref}}[G]-\text{const.},
\label{eq:Phi_def_app}
\end{equation}
so that
\begin{equation}
\Gamma[G]
=
\Gamma_{\mathrm{ref}}[G]+\Phi[G]+\text{const.}
\label{eq:Gamma_split_app}
\end{equation}
Differentiating gives
\begin{equation}
\frac{\delta \Gamma_{\mathrm{ref}}[G]}{\delta G}
=
G^{-1}-G_0^{-1},
\qquad
\Sigma[G]
=
\frac{\delta \Phi[G]}{\delta G},
\label{eq:dGamma_ref_and_Sigma_app}
\end{equation}
hence at stationarity
\begin{equation}
G^{-1}
=
G_0^{-1}
-
\Sigma[G].
\label{eq:Dyson_from_Phi_app}
\end{equation}
Differentiating once more yields the exact irreducible two-particle kernel
\begin{equation}
K_{\mathrm{irr}}
=
\frac{\delta^2 \Phi[G]}{\delta G\,\delta G}.
\label{eq:Kirr_def_app}
\end{equation}

\subsection{The retained interaction part and the RPA kernel split}

For the direct RPA-type closure, one chooses the retained interaction kernel in the direct particle-hole channel to be the bare Coulomb kernel and writes
\begin{equation}
K_{\mathrm{irr}} = v + K_{\mathrm{rem}},
\label{eq:Kirr_split_app}
\end{equation}
where \(K_{\mathrm{rem}}\) denotes the remaining irreducible contribution. This is the kernel-level split used in the main text.\cite{Baym1962,LuttingerWard1960,Klein1961} It need not arise from a unique global decomposition of the full functional \(\Phi[G]\), but it is the natural split for the direct RPA closure.\cite{Potthoff2006,KozikFerreroGeorges2015,Vucicevic2018}

\subsection{Equal-time reduction and density-channel contraction}

Let \(x=(\mathbf r,\sigma)\) and \(1=(x,\tau)\). Define the equal-time reduction map \(\Pet\) on bilocal kernels by
\begin{equation}
(\Pet G)(x,x';\tau)
:=
\lim_{\eta\downarrow 0}
G\bigl((x,\tau),(x',\tau+\eta)\bigr),
\label{eq:Pet_def_app}
\end{equation}
whenever the one-sided limit exists. In equilibrium, the right-hand side is \(\tau\)-independent, and one identifies the one-body reduced density matrix as
\begin{equation}
\gamma(x,x') = (\Pet G)(x,x';\tau).
\label{eq:gamma_from_Pet_app}
\end{equation}
Next define the diagonal density map \(\Pdiag\) on equal-time bilocal kernels by
\begin{equation}
(\Pdiag \gamma)(x) := \gamma(x,x).
\label{eq:Pdiag_def_app}
\end{equation}
Thus
\begin{equation}
\rho(x)
=
(\Pdiag\gamma)(x)
=
\lim_{\eta\downarrow 0}
G\bigl((x,\tau),(x,\tau+\eta)\bigr).
\label{eq:rho_from_limits_app}
\end{equation}

To contract response kernels, one also needs the corresponding source embeddings. Let \(u(x',x)\) be a static equal-time bilocal source. Its embedding into the full spacetime-bilocal source space is
\begin{equation}
(\Iet u)(2,1)
:=
u(x_2,x_1)\,\delta(\tau_2-\tau_1),
\label{eq:Iet_def_app}
\end{equation}
and if \(j(x,\tau)\) is a local density source, its embedding is
\begin{equation}
(\Idiag j)(2,1)
:=
j(x_1,\tau_1)\,\delta(x_2-x_1)\,\delta(\tau_2-\tau_1).
\label{eq:Idiag_def_app}
\end{equation}
With these definitions,
\begin{equation}
\Lambda = \Pet\,L\,\Iet,
\qquad
\chi = \Pdiag\,\Pet\,L\,\Idiag.
\label{eq:response_contractions_app}
\end{equation}
The first contraction uses the static source embedding and therefore selects the zero-transfer sector appropriate to the static 1RDM response. The density-source embedding may carry arbitrary bosonic transfer frequency, and its contraction gives the dynamical density response evaluated about the equilibrium background.

\section{Grand-canonical finite-temperature formulations of static DFT and 1RDMFT}
\label{app:grandcanonical_DFT_1RDMFT}

This appendix records the grand-canonical finite-temperature source formulations corresponding to the static density and equal-time 1RDM levels. It also states explicitly the sector-restricted finite-temperature constrained searches whose zero-temperature limits give the fixed-\(N\) ensemble functionals used in Sections~\ref{subsec:dftlevel} and \ref{subsec:1rdmlevel}. The dynamical density source formulation is treated directly in Section~\ref{subsec:lrtdftlevel}, since it is the natural entry point for LR-TDDFT.

\subsection{Density level}

For a fixed coupling constant \(\lambda\in[0,1]\), define the
grand-canonical operator
\[
\hat K^\lambda[v,\mu]
=
\hat T+\lambda\hat W+\hat V[v]-\mu\hat N,
\qquad
\hat V[v]=\int dr\,v(r)\hat\rho(r).
\]
The grand potential is
\[
\Omega^\lambda_{\beta,\mu}[v]
=
-\frac{1}{\beta}
\ln
\operatorname{Tr}_{\mathcal F}
e^{-\beta\hat K^\lambda[v,\mu]} .
\tag{E1}
\]
By the Gibbs variational principle,
\[
\Omega^\lambda_{\beta,\mu}[v]
=
\inf_\Gamma
\left\{
\operatorname{Tr}_{\mathcal F}\Gamma\hat K^\lambda[v,\mu]
+
\frac{1}{\beta}
\operatorname{Tr}_{\mathcal F}(\Gamma\ln\Gamma)
\right\},
\tag{E2}
\label{eq:Gibbs_density_app}
\]
where the infimum is over grand-canonical density matrices
on fermionic Fock space.

Grouping the trial states according to the density they generate
leads to the intrinsic finite-temperature density functional
\[
A^\lambda_\beta[\rho]
=
\inf_{\Gamma\mapsto\rho}
\left\{
\operatorname{Tr}_{\mathcal F}\Gamma(\hat T+\lambda\hat W)
+
\frac{1}{\beta}
\operatorname{Tr}_{\mathcal F}(\Gamma\ln\Gamma)
\right\}.
\tag{E3}
\]
Since \(\hat V[v]-\mu\hat N\) couples to the density through
the scalar source \(v-\mu\), Eq.~\eqref{eq:Gibbs_density_app} becomes
\[
\Omega^\lambda_{\beta,\mu}[v]
=
\inf_\rho
\left\{
A^\lambda_\beta[\rho]
+
\int dr\, [v(r)-\mu]\rho(r)
\right\}.
\tag{E4}
\]
This is the Mermin finite-temperature density-functional
variational principle.\cite{Mermin1965}

For a fixed external potential \(v\), let \(E_v^\lambda(M)\) denote the ground-state energy of \(\hat T+\lambda\hat W+\hat V[v]\) in the \(M\)-particle sector. If \(\mu\) lies in the corresponding addition--removal interval,
\[
E_v^\lambda(N)-E_v^\lambda(N-1)
\le
\mu
\le
E_v^\lambda(N+1)-E_v^\lambda(N),
\tag{E5}
\]
then the equilibrium grand-canonical state approaches an \(N\)-particle ground state as \(\beta\to\infty\). This equilibrium sector-selection statement should be distinguished from a pointwise limit of the intrinsic Fock-space constrained search, for which fixing \(\int\rho=N\) fixes only the mean particle number.

To connect directly with the fixed-\(N\) functional used in the main text, define the sector-restricted finite-temperature constrained search
\[
\begin{aligned}
A^\lambda_{\beta,N}[\rho]
={}&
\inf_{\substack{\Gamma_N\in\mathcal D(\mathcal H_N)\\
                  \Gamma_N\mapsto\rho}}
\Bigl\{
\operatorname{Tr}_{\mathcal H_N}
\Gamma_N(\hat T+\lambda\hat W)
\\[-1mm]
&\hspace{28mm}
+
\beta^{-1}
\operatorname{Tr}_{\mathcal H_N}(\Gamma_N\ln\Gamma_N)
\Bigr\}.
\end{aligned}
\tag{E6}
\]
In a finite-dimensional regularization, and more generally under the appropriate compactness and lower-semicontinuity assumptions,
\[
\begin{aligned}
A^\lambda_{\beta,N}[\rho]
&\longrightarrow
F_N^\lambda[\rho],
\qquad \beta\to\infty,
\\
F_N^\lambda[\rho]
&:=
\inf_{\substack{\Gamma_N\in\mathcal D(\mathcal H_N)\\
                  \Gamma_N\mapsto\rho}}
\operatorname{Tr}_{\mathcal H_N}
\Gamma_N(\hat T+\lambda\hat W).
\end{aligned}
\tag{E7}
\]

\subsection{Equal-time bilocal level}

For a fixed coupling constant \(\lambda\in[0,1]\), the 1RDM
formulation is parallel, with the local scalar source replaced by
a Hermitian equal-time bilocal one-body source. Let
\[
\hat U[u]
=
\iint drdr'\,
u(r',r)\hat\psi^\dagger(r')\hat\psi(r),
\]
and define
\[
\hat K^\lambda[u,\mu]
=
\hat T+\lambda\hat W+\hat U[u]-\mu\hat N .
\]
The grand potential is
\[
\Omega^\lambda_{\beta,\mu}[u]
=
-\frac{1}{\beta}
\ln
\operatorname{Tr}_{\mathcal F}
e^{-\beta\hat K^\lambda[u,\mu]} .
\tag{E8}
\]
Again, the Gibbs variational principle gives
\[
\Omega^\lambda_{\beta,\mu}[u]
=
\inf_\Gamma
\left\{
\operatorname{Tr}_{\mathcal F}\Gamma\hat K^\lambda[u,\mu]
+
\frac{1}{\beta}
\operatorname{Tr}_{\mathcal F}(\Gamma\ln\Gamma)
\right\}.
\tag{E9}
\]

Grouping the trial states according to the 1RDM they generate
defines the intrinsic finite-temperature 1RDM functional~\cite{Baldsiefen2015,SutterGiesbertz2023}
\[
A^\lambda_\beta[\gamma]
=
\inf_{\Gamma\mapsto\gamma}
\left\{
\operatorname{Tr}_{\mathcal F}\Gamma(\hat T+\lambda\hat W)
+
\frac{1}{\beta}
\operatorname{Tr}_{\mathcal F}(\Gamma\ln\Gamma)
\right\}.
\tag{E10}
\]
Since \(\hat U[u]\) couples linearly to \(\gamma\), while
\(\hat N\) contributes \(\operatorname{Tr}\gamma\), one obtains
\[
\Omega^\lambda_{\beta,\mu}[u]
=
\inf_\gamma
\left\{
A^\lambda_\beta[\gamma]
+
\iint drdr'\,u(r',r)\gamma(r,r')
-
\mu\operatorname{Tr}\gamma
\right\}.
\tag{E11}
\]
Equivalently, the chemical-potential term may be absorbed into
the bilocal source,
\[
u_\mu(r',r)
=
u(r',r)-\mu\delta(r-r').
\tag{E12}
\]

For a fixed bilocal source \(u\), a chemical potential in the corresponding addition--removal interval similarly selects the \(N\)-particle equilibrium sector as \(\beta\to\infty\). To connect pointwise with the fixed-\(N\) 1RDM functional of the main text, define instead
\[
\begin{aligned}
A^\lambda_{\beta,N}[\gamma]
={}&
\inf_{\substack{\Gamma_N\in\mathcal D(\mathcal H_N)\\
                  \Gamma_N\mapsto\gamma}}
\Bigl\{
\operatorname{Tr}_{\mathcal H_N}
\Gamma_N(\hat T+\lambda\hat W)
\\[-1mm]
&\hspace{28mm}
+
\beta^{-1}
\operatorname{Tr}_{\mathcal H_N}(\Gamma_N\ln\Gamma_N)
\Bigr\}.
\end{aligned}
\tag{E13}
\]
Under the same regularization or compactness assumptions,
\[
\begin{aligned}
A^\lambda_{\beta,N}[\gamma]
&\longrightarrow
F_N^\lambda[\gamma],
\qquad \beta\to\infty,
\\
F_N^\lambda[\gamma]
&:=
\inf_{\substack{\Gamma_N\in\mathcal D(\mathcal H_N)\\
                  \Gamma_N\mapsto\gamma}}
\operatorname{Tr}_{\mathcal H_N}
\Gamma_N(\hat T+\lambda\hat W).
\end{aligned}
\tag{E14}
\]
The physical zero-temperature functional is recovered at
\(\lambda=1\), namely \(F_N[\gamma]=F_N^1[\gamma]\); the subscript \(N\) is suppressed in the main text.

\subsection{Relation to the main text}

The main text uses the fixed-\(N\) zero-temperature ensemble functionals \(F_N[\rho]\) and \(F_N[\gamma]\), with the particle-number subscript suppressed, because they expose the static density and equal-time bilocal Hessian structures most directly. The grand-canonical principles above provide the natural finite-temperature source formulations and select the equilibrium particle-number sector for an appropriate chemical potential. Their pointwise connection to the canonical functionals of the main text is made by the explicit \(N\)-sector restrictions in Eqs.~(E6) and (E13), whose zero-temperature limits are Eqs.~(E7) and (E14).

\bibliography{ref}

\end{document}